\documentclass[12pt]{article}
\def\cdate{{August 1, 2019}}
\usepackage{amsfonts}
\usepackage{amsmath}
\usepackage{latexsym}
\usepackage{amssymb}
\usepackage{txfonts}
\usepackage[american]{babel}

\makeatletter
\def\timenow{%
\@tempcnta=\time \divide\@tempcnta by 60 \number\@tempcnta:\multiply
\@tempcnta by 60 \@tempcntb=\time \advance\@tempcntb by -\@tempcnta
\ifnum\@tempcntb <10 0\number\@tempcntb\else\number\@tempcntb\fi}
\newcounter{outputpage}
\renewcommand{\@oddhead}{\stepcounter{outputpage}\hfill\hfill\theoutputpage}
\renewcommand{\@evenhead}{\stepcounter{outputpage}\hfill\hfill\theoutputpage}
\renewcommand{\@oddfoot}{\vbox{\hrule\vspace{3pt}\hfil{\scriptsize\textit{
\hfill\hfill\jobname.tex; \today; \timenow; p. \theoutputpage}}\hfil}}
\renewcommand{\@evenfoot}
{\vbox{\hrule\vspace{3pt}\hfil{\scriptsize\textit{
-
\hfill\hfill\jobname.tex; \today; \timenow; p. \theoutputpage}}\hfil}}
\makeatother

\def\RR{{\mathbb R}}

\def\cA{{\cal A}}
\def\cB{{\cal B}}
\def\cC{{\cal C}}
\def\cD{{\cal D}}
\def\cE{{\cal E}}
\def\cH{{\cal H}}

\def\cM{{\cal M}}
\def\cF{{\cal F}}
\def\cV{{\cal V}}
\def\cW{{\cal W}}
\def\cP{{\cal P}}
 
\def\cR{\mathcal{R}}

\def\tr{\mathrm{ tr}} 
\def\Tr{\mathrm{ Tr}}

\def\End{\mathrm{End}}

\def\be{\begin{equation}} 
\def\ee{\end{equation}} 
\def\bea{\begin{eqnarray}} 
\def\eea{\end{eqnarray}} 
\def\bed{\begin{definition}{\ }}
\def\eed{\end{definition}}
\def\bd{\begin{description}}
\def\ed{\end{description}}
\def\bc{\begin{center}}
\def\ec{\end{center}}
\newtheorem{theorem}{Theorem}
\newtheorem{lemma}{Lemma}
\newtheorem{corollary}{Corollary}

\newtheorem{definition}{Definition}
\def\sideremark#1{\ifvmode\leavevmode\fi\vadjust{\vbox to0pt{\vss
\hbox to 0pt{\hskip\hsize\hskip1em
\vbox{\hsize2cm\tiny\raggedright\pretolerance10000
\noindent #1\hfill}\hss}\vbox to8pt{\vfil}\vss}}}

\begin{document}

\begin{titlepage}

\null
\vspace{-10mm}
\hspace*{50truemm}{\hrulefill}\par\vskip-4truemm\par
\hspace*{50truemm}{\hrulefill}\par\vskip5mm\par
\hspace*{50truemm}{{\large\sc New Mexico Tech {\rm 
(\cdate)}}}\vskip4mm\par
\hspace*{50truemm}{\hrulefill}\par\vskip-4truemm\par
\hspace*{50truemm}{\hrulefill}
\par
\bigskip
\bigskip
\par
\par
\vfill
\centerline{\huge\bf Relative Spectral Invariants}
\bigskip
\centerline{\huge\bf of Elliptic Operators on Manifolds}
\bigskip
\bigskip
\centerline{\Large\bf Ivan G. Avramidi
}
\bigskip
\centerline{\it Department of Mathematics}
\centerline{\it New Mexico Institute of Mining and Technology}
\centerline{\it Socorro, NM 87801, USA}
\centerline{\it E-mail: ivan.avramidi@nmt.edu}
\bigskip
\medskip
\vfill
{\narrower
\par

We introduce and study {\it new} relative spectral invariants of {\it 
two} elliptic partial differential operators of Laplace and Dirac type on 
compact smooth manifolds without boundary that depend on both the 
eigenvalues and the eigensections of these operators
and contain much more information about geometry. 
We prove the existence of the homogeneous short time asymptotics of the new
invariants with the coefficients of the asymptotic expansion being
integrals of some invariants that depend on the symbols of both operators.
The first two coefficients of the asymptotic expansion are computed explicitly.


\par}
\vfill
\vfill
{\vbox{
\hrule
\vspace{3pt}
{\scriptsize{\it
\hfill\hfill 
Ivan G. Avramidi; 
\jobname.tex; 
\today; 
\timenow}}}}

\end{titlepage}

 
\section{Introduction}
\setcounter{equation}0


Elliptic operators on manifolds, in particular, first and second-order partial 
differential operators, play a crucial role in global analysis, spectral 
geometry and mathematical physics 
\cite{gilkey95,berline92,berger03,avramidi00,avramidi10,avramidi15}. The study 
of the spectrum of elliptic operators is of paramount importance since it 
describes various important objects in quantum field theory and differential 
geometry such as correlation functions, functional determinants, integrals of 
infinite-dimensional Hamiltonian systems etc. The spectrum of elliptic 
operators does, of course, depend on the geometry of the manifold. Therefore, 
one can ask the question: ``To what extent does the spectrum of a {\it single} 
elliptic operator describe the geometry?'', or, as M. Kac put it ``Can one hear 
the shape of a drum?'' In general one cannot compute the spectrum exactly. One 
usually studies the spectrum indirectly by studying some spectral invariants 
such as the heat trace or zeta function \cite{gilkey95}. These spectral 
invariants only depend on the eigenvalues of the operators and do not depend on 
the eigensections. It is well known now that the answer to this question is 
negative, that is, there are non-isometric manifolds that have the same 
spectrum. The classical heat trace of Laplace type operators has been studied 
for decades going back to H. Weyl \cite{weyl11} and S. Minakshisundaram and A. 
Plejel \cite{minakshisundaram49}. There is a vast literature on the subject 
(see \cite{gilkey95,berline92,avramidi00,avramidi10,avramidi15} and references 
therein). 

In \cite{avramidi16} and 
\cite{avramidi17} we studied more general spectral invariants that appear 
naturally in quantum statistical physics and geometry. In the present paper
we introduce and study {\it new relative spectral invariants} of {\it 
two} elliptic operators. We hope that these new invariants could shed new light 
on the old questions of spectral geometry. We generalize the question as 
follows: ``Does the spectral data  of {\it two} elliptic operators determine 
the 
geometry?'' These invariants depend both on the eigenvalues and the 
eigensections and contain much more information about geometry. 
Such relative spectral invariants appear naturally, in 
particular, in the study of particle creation in quantum field theory and 
quantum gravity \cite{birrel80,dewitt75,avramidi19a}. They determine the 
number 
of created particles from the vacuum when the dynamical operator depends on 
time. 

In Sec. 2 we motivate the study of the relative spectral invariants.
We describe the so-called {\it Bogolyubov invariant} in quantum field theory
and show how it can be expressed in terms of the relative spectral invariant.
We consider a smooth $n$-dimensional compact manifold $M$ without boundary
and a vector bundle $\cV$ over the 
manifold $M$. Let $L_\pm$ be two self-adjoint elliptic second-order partial 
differential operators acting on smooth sections of the vector bundle $\cV$
with a positive definite scalar leading symbols of Laplace type. 
Let $D_\pm$ be two self-adjoint elliptic first-order partial differential 
operators acting on smooth sections of the vector bundle $\cV$ of
Dirac type such that the squares $L_\pm=D_\pm^2$ are Laplace type operators.
The spectral information about the operators $L_\pm$ and $D_\pm$
are contained in the classical heat traces
\bea
\Theta_\pm(t) &=& \Tr \exp(-tL_\pm),
\label{237ssb}
\\
H_\pm(t) &=& \Tr D_\pm \exp(-tD_\pm^2).
\label{12via}
\eea
We show that the Bogolyubov invariants can be expressed in terms of the
traces
\bea
\Psi(t,s) &=&
\Tr\left\{\exp(-tL_+)-\exp(-tL_-)\right\}
\left\{\exp(-sL_+)-\exp(-sL_-)\right\},
\label{220ssb}
\\
\Phi(t,s) &=&
\Tr\left\{D_+\exp(-tD^2_+)-D_-\exp(-tD^2_-)\right\}
\left\{D_+\exp(-sD^2_+)-D_-\exp(-sD^2_-)\right\},
\nonumber\\
\label{248ssb}
\eea
that we call {\it relative spectral invariants};
which can further be expressed
further in terms of the classical heat traces
and the {\it combined heat traces}
\bea
X(t,s) &=& \Tr\exp(-tL_+)\exp(-s L_-),
\label{238ccx}
\\
Y(t,s) &=& \Tr D_+\exp(-tD^2_+)D_-\exp(-sD^2_-),
\label{18ssb}
\eea
by
\bea
\Psi(t,s) &=& \Theta_+(t+s)+\Theta_-(t+s)
-X(t,s)-X(s,t),
\label{16via}
\\
\Phi(t,s) &=& -\partial_t\Theta_+(t+s)-\partial_t\Theta_-(t+s)
-Y(t,s)-Y(s,t).
\label{17via}
\eea

In Sec. 3 we describe the relevant differential operators and 
their spectral traces and introduce the relevant notation.
The operators $L_\pm$ naturally define the metrics $g_{ij}^\pm$,
the connection one forms $\cA^\pm_i$ and the endomorphisms $Q_\pm$ by 
\be
L_\pm = -g_\pm^{-1/4}(\partial_i+\cA^\pm_i)g_\pm^{1/2}g_\pm^{ij}
(\partial_j+\cA^\pm_j)g_\pm^{-1/4}+Q_\pm,
\ee
where $g^{ij}_\pm$ are the inverse metrics and $g_\pm=\det g^\pm_{ij}$. 
Similarly, the operators $D_\pm$ define the endomorphisms $S_\pm$ by
\be
D_\pm = g_\pm^{1/4}i\gamma^j_\pm(\partial_j+\cA_j^\pm)g_\pm^{-1/4}+S_\pm,
\ee
where $\gamma_\pm^j$ are the Dirac matrices satisfying (\ref{39viax}),
here the connection $\cA^\pm_i$ is supposed to satisfy the compatibility
condition (\ref{310viax}). Also, we suppose that the 
endomorphisms $S_\pm$ anticommute with the
Dirac matrices $\gamma_\pm^i$,
 (\ref{312viax}),
so that the square of the Dirac type
operator  $D_\pm^2$ is a Laplace type operator 
with the potential
\be
Q_\pm=-\frac{1}{2}\gamma^{ij}_\pm\cR^\pm_{ij}
+S_\pm^2+i\gamma_\pm^j\nabla^\pm_j S_\pm,
\label{qsxxab}
\ee
where 
$\cR^\pm_{ij}$ be the curvature of the connection $\cA^\pm_i$
and $\gamma_\pm^{ij}=\gamma_\pm^{[i}\gamma_\pm^{j]}$.
We follow the standard convention \cite{zhelnorovich19}
and denote the
antisymmetrized products of Dirac matrices by
$\gamma^{i_1\dots i_k}=\gamma^{[i_1}\cdots\gamma^{i_k]}$.

In Sec. 4 we present a detailed review of the Ruse-Synge function
(which is equal to one half of the square of the geodesic distance
between two points in a Riemannian manifold)
with the particular emphasis on its dependence on the metric. 
We compute the diagonal values of the covariant derivatives
(defined with respect to a metric $g$) of a Ruse-Synge function
$\sigma^h(x,x')$ defined with respect to another metric $h$.

In Sec. 5 we study the asymptotics of the integrals of Laplace type
and prove some important lemmas used in the proof of the main theorems.
In Sec. 6 we study the asymptotics of the combined heat traces
and prove the general theorems.
The spectral information about the operators $L_\pm$
is contained in the
classical heat traces (\ref{237ssb}).
In particular, the asymptotic expansion as $t\to 0$ 
\bea
\Theta_\pm\left(t\right) 
&\sim& (4\pi)^{-n/2}
\sum_{k=0}^\infty t^{k-n/2}A^\pm_k,
\label{513xxc}
\eea
defines the sequence of spectral invariants
\bea
A^\pm_k &=& \frac{(-1)^k}{k!}\int\limits_M dx\; g^{1/2}_\pm \tr[a^\pm_k],
\eea
where $\tr[a^\pm_k]$ are
some scalar invariants
which are polynomial in the jets of the symbols of the operators
$L_\pm$, that is, in the covariant derivatives of the
curvatures $R^\pm_{ijkl}$
of the metrics $g_\pm$, the curvatures $\cR^\pm_{ij}$ of the
connections $\cA^\pm_i$ and the potentials $Q_\pm$
(notice the different normalization factor in (\ref{513xxc})
 compared to our earlier
papers
\cite{avramidi91,avramidi00,avramidi10,avramidi15}).
It is well known that the first two classical heat kernel coefficients are
\cite{gilkey95,avramidi00,avramidi15}
\bea
A^\pm_0 &=& \int_M dx\;g_\pm^{1/2} \tr I,
\\{}
A^\pm_1 &=& \int_M dx\;g^{1/2}_\pm \tr\left(\frac{1}{6}R_\pm I-Q_\pm\right),
\eea
where $\tr$ is the fiber trace,
$I$ is the identity endomorphism and
$R_\pm$ is the scalar curvature of the metric $g_\pm$.
Therefore, 
for the Dirac type operators the coefficient $A_1$ takes the form
\bea
A^\pm_1 &=& \int_M dx\;g^{1/2}_\pm \tr\left(\frac{1}{6}R_\pm I
+\frac{1}{2}\gamma^{ij}_\pm\cR^\pm_{ij}-S_\pm^2
\right).
\label{126viax}
\eea

We study in this paper the asymptotics of the 
he combined heat traces (\ref{238ccx}) and (\ref{18ssb}).
We define the time-dependent
metric $g_{ij}=g_{ij}(t,s)$ as the inverse of the matrix
\be
g^{ij}=tg_+^{ij}+sg_-^{ij},
\label{113via}
\ee
with $t,s> 0$; throughout the paper we use the notation
$g=\det g_{ij}$ for the determinant of the metric. 
Also, we define the time-dependent connection 
$\cA_i=\cA_i(t,s)$
by
\be
\cA_i=g_{ij}\left(tg^{jk}_+\cA^+_k+sg^{jk}_-\cA^-_k\right).
\label{114via}
\ee
and the vectors
\be
\cC^\pm_i = \cA^\pm_i-\cA_i.
\label{115via}
\ee
We omit the variables $t$ and $s$ where it does not
cause any confusion.
Note that the inverse metric $g^{ij}$ is a homogeneous function of $t$ and 
$s$ of degree $1$, and, therefore, the metric $g_{ij}$ is a homogeneous 
function of $t$ and $s$ of degree $(-1)$, and the determinant $g=\det g_{ij}$  
is a homogeneous function of $t$ and $s$ of degree $(-n)$; furthermore, the 
Christoffel symbols, $\Gamma_{g}{}^i{}_{jk}$, the Riemann tensor 
$R_g{}^i{}_{jkl}$ and the Ricci tensor $R^g_{ij}$ of the metric $g$ are 
homogeneous functions of $t$ and $s$ of degree $0$. Similarly, the connection 
$\cA_i$ and its curvature $\cR^\cA_{ij}$ are  homogeneous functions of $t$ and $s$ 
of degree $0$.

\begin{theorem}
\label{theorem1}
There are asymptotic expansions as $\varepsilon\to 0$
\bea
X(\varepsilon{}t,\varepsilon{}s) &\sim& (4\pi\varepsilon)^{-n/2}
\sum_{k=0}^\infty \varepsilon^{k} B_k(t,s),
\label{1zaab}
\\
Y(\varepsilon{}t,\varepsilon{}s) &\sim&
(4\pi\varepsilon)^{-n/2}
\sum_{k=0}^\infty \varepsilon^{k-1} 
C_k(t,s),
\label{15zaac}
\eea
where 
\bea
B_k(t,s) &=& \int\limits_M dx\; g^{1/2}(t,s)b_k(t,s),
\\
C_k(t,s) &=& \int\limits_M dx\; g^{1/2}(t,s)c_k(t,s).
\eea
\begin{enumerate}
\item 
The coefficients $b_k(t,s)$ and $c_k(t,s)$
are scalar invariants built polynomially from the covariant
derivatives (defined with respect to the metric $g_{ij}$ and the
connection $\cA_i$)
of the metrics $g^\pm_{ij}$, the vectors $\cC^\pm_i$ and
the potentials $Q_\pm$ and $S_\pm$.

\item
The coefficients $b_k(t,s)$ and $c_k(t,s)$
are symmetric under the exchange $(t,L_+)\leftrightarrow (s,L_-)$.

\item
The coefficients $b_k(t,s)$
are homogeneous functions of $t$ and $s$
of degree $k$ and the coefficients $c_k(t,s)$ are 
homogeneous
functions of $t$ and $s$
of degree $(k-1)$.
\end{enumerate}
\end{theorem}

This gives the asymptotic expansion of the relative spectral
invariants (\ref{220ssb}) and (\ref{248ssb}).

\begin{corollary}
\label{corollary1}
There are asymptotic expansions as $\varepsilon\to 0$
\bea
\Psi(\varepsilon{}t,\varepsilon{}s) 
&\sim& (4\pi\varepsilon)^{-n/2}
\sum_{m=0}^\infty \varepsilon^{k} \Psi_k(t,s),
\label{120via}
\\
\Phi(\varepsilon{}t,\varepsilon{}s) 
&\sim&
(4\pi\varepsilon)^{-n/2}
\sum_{k=0}^\infty \varepsilon^{k-1} 
\Phi_k(t,s),
\label{121via}
\eea
where
\bea
\Psi_k(t,s) &=&
(t+s)^{k-n/2}(A_k^++A_k^-)
-B_k(t,s)-B_k(s,t),
\label{543xxca}
\\
\Phi_k(t,s) &=&
-\left(k-\frac{n}{2}\right)(t+s)^{k-1-n/2}\left(A_k^++A_k^-\right)
-C_k(t,s)-C_k(s,t).
\label{542saax}
\eea

\end{corollary}

In Sec. 7 we consider some particular cases when the relative spectral
invariants can be computed exactly in terms of the classical heat trace
and compute explicitly the first two coefficients of the asymptotic expansions.
To describe the main results we introduce 
a symmetric tensor $G_{ij}=G_{ij}(t,s)$ 
(that we call the {\it dual metric})
by
\be
G_{ij} = s g^+_{ij} + t g^-_{ij},
\label{126via}
\ee
and its inverse $G^{ij}$,
which is related to the metric $g^{ij}$, (\ref{113via}), 
by
\bea
G_{ij} & =& g^+_{ik}g^{kl}g^-_{lj}
= g^-_{ik}g^{kl}g^+_{lj},
\label{128viab}
\\
G^{ij} &=&  g_-^{ip}g_{pq}g_+^{qj}
=  g_+^{ip}g_{pq}g_-^{qj}.
\label{127via}
\eea
Notice that
\bea
g_{ij}(1,0) &=& G_{ij}(0,1)=g^+_{ij},
\\
g_{ij}(0,1) &=& G_{ij}(1,0)=g^-_{ij}.
\eea

Also, we introduce the non-compatibility tensors
\bea
K^\pm_{ijk} &=& \nabla_i^g g^\pm_{jk},
\label{128via}
\eea
and the tensors
\bea
W_\pm{}^i{}_{jk} &=& 
\frac{1}{2}g_\pm^{im}\left(K^\pm_{jkm}+K^\pm_{kjm}
-K^\pm_{mjk}\right),
\label{129via}
\\
W^\pm_j &=& W_\pm{}^i{}_{ij}
=\nabla^g_j W^\pm,
\label{130via}
\eea
with
\be
W^\pm = \frac{1}{2}\log\left(\frac{g_\pm}{g}\right).
\label{131via}
\ee
Finally, we define
\bea
W_i &=& \frac{1}{2}\left(W^+_i+W^-_i\right)
=\frac{1}{2}\nabla^g_i\left(W^++W^-\right),
\label{137zax}
\\
W_{ij}&=& \frac{1}{2}\left(\nabla^g_jW^+_i
+\nabla^g_jW^-_i\right)
=\frac{1}{2}\nabla^g_i\nabla^g_j(W^++W^-),
\label{130viaxz}
\\
\Sigma_{ijk} 
&=& \frac{3}{2}sK^+_{(ijk)}+\frac{3}{2}tK^-_{(ijk)},
\label{139zax}
\\
\Sigma_{ijkl}
&=& s S^+_{ijkl} 
+ t S^-_{ijkl},
\label{140zax}
\eea
where
\be
S^\pm_{ijkl} =
4g^\pm_{m(i} \nabla^g{}_{j}W_\pm^m{}_{kl)}
+4g^\pm_{m(i}W_\pm^n{}_{jk}W_\pm^{m}{}_{l)n}
+3g^{\pm}W_\pm^n{}_{(ij}W_\pm^{m}{}_{kl)}.
\label{141zax}
\ee
Here and everywhere below parenthesis denote symmetrization over all
indices included. The indices excluded from the symmetrization are separated by
vertical lines.


\begin{theorem}
\label{theorem2}
The first two coefficients of the asymptotic expansion of the combined
heat trace $X(t,s)$ are
\bea
b_0(t,s) &=& \tr I,
\label{132via}
\\
b_1(t,s) &=&
\tr\Biggl\{
t\left(\frac{1}{6}R_+I
-Q_+\right)
+s\left(\frac{1}{6}R_-I
-Q_-\right)
+ts\Biggl[\frac{1}{6}G^{ij}\left(
R^+_{ij}+R^-_{ij}
-2R^g_{ij}\right)I
\nonumber\\
&&
+\Biggl(\frac{1}{6}
G^{ij}\left(W_{ij}
+W_{i}W_{j}\right) 
-G^{ij}G^{kl}\Sigma_{ikl}W_{j}
-\frac{1}{4}G^{ij}G^{kl}\Sigma_{ijkl}
\nonumber\\
&&
+\frac{1}{12}\left(2G^{il}G^{jm}
+3G^{ij}G^{lm}\right)G^{kn}\Sigma_{ijk}\Sigma_{lmn}
\Biggr)I
\nonumber\\
&&
+G^{ij}(\cC^+_{i}-\cC^-_{i})(\cC^+_{j}-\cC^-_{j})
\Biggr]
\Biggr\}.
\label{134via}
\eea
\end{theorem}

\begin{corollary}
\label{corollary2}
The first two coefficients of the asymptotic expansion of the
relative spectral invariant $\Psi(t,s)$ are
\bea
\Psi_0(t,s) &=& \int\limits_M dx\; 
\left\{(t+s)^{-n/2}\left(g_+^{1/2}+g_-^{1/2}\right)
-g^{1/2}(t,s)-g^{1/2}(s,t)\right\}\tr I,
\\
\Psi_1(t,s) &=& \int_M dx\;\Biggl\{
(t+s)^{1-n/2}\Biggl[g_+^{1/2}\tr\left(\frac{1}{6}R_+ I-Q_+\right)
+g_-^{1/2}\tr\left(\frac{1}{6}R_-I-Q_-\right)\Biggr]
\nonumber\\
&&
-g^{1/2}(t,s)b_1(t,s)-g^{1/2}(s,t)b_1(s,t)
\Biggr\}.
\label{543xxcab}
\eea
\end{corollary}

Further, we define the auxiliary tensors
\bea
N^{jkl} &=&
2G^{ij}G^{kl}W_{i}
-\frac{1}{3}\left(
2G^{ij}G^{qk}
+3G^{iq}G^{jk}
\right)G^{pl}\Sigma_{ipq},
\label{144viax}
\\
M^{kl} &=& 
\left(G^{kl}G^{ij}
+2G^{ik}G^{jl}\right)
(W_{ij}+W_iW_j)
\nonumber\\
&&
-\left(2G^{ij}G^{mk}G^{pl}
+2G^{im}G^{jk}G^{pl}
+G^{kl}G^{im}G^{pj}
\right)\Sigma_{pim}W_{j}
\nonumber\\
&&
-\frac{1}{4}\left(G^{pq}G^{kl}
+4G^{kp}G^{lq}\right)G^{pq}\Sigma_{ijpq}
\nonumber\\
&&
+\frac{1}{72}\Biggl(2G^{ij}G^{pr}G^{qs}G^{kl}
+3G^{ij}G^{pq}G^{rs}G^{kl}
+6G^{ik}G^{jl}G^{pq}G^{rs}
\nonumber\\
&&
+12G^{ij}G^{pq}G^{kr}G^{ls}
+12G^{ij}G^{pr}G^{kq}G^{sl}
\Biggr)\Sigma_{ipq}\Sigma_{jrs}.
\label{147zax}
\eea
and
\bea
V_{pqijkl}&=& \mathrm{Sym}(i,j,k,l)
\Biggl\{
\Biggl(
4g^+_{mp} \nabla^g{}_{k}W_+^m{}_{ij}
+4g^+_{mp}W_+^n{}_{jk}W_+^{m}{}_{in}
+12g^+_{mi}W_+{}^m{}_{nk}W_+{}^n{}_{jp}
\nonumber\\
&&
-6g^+_{mi}W_+{}^n{}_{kj}W_+{}^m{}_{np}
\Biggr)
g^-_{lq}
+g^+_{lp}\Biggl(
4g^-_{mq} \nabla^g{}_{k}W_-^m{}_{ij}
+4g^-_{mq}W_-^n{}_{jk}W_-^{m}{}_{in}
\nonumber\\
&&
+12g^-_{mi}W_-{}^m{}_{nk}W_-{}^n{}_{jq}
-6g^-_{mi}W_-{}^n{}_{kj}W_-{}^m{}_{nq}
\Biggr)
+6g^+_{mp}W_+^m{}_{ij}g^-_{nq}W_-^n{}_{kl}
\Biggr\}.
\label{145viaxz}
\eea

\begin{theorem}
\label{theorem3}
The first two coefficients of the asymptotic expansion of the combined
heat trace $Y(t,s)$ are
\bea
c_0(t,s) &=&
\frac{1}{2}g_{ij}(t,s)\tr\,
\left(\gamma_+^{i}\gamma_-^{j}\right),
\label{138via}
\\
c_1(t,s) &=&
\tr\Biggl\{
\frac{1}{6}t\left(\frac{1}{2}g_{pq} R_+
-g_{qi}g_+^{ij}R^+_{jp} 
\right) \gamma_+^p\gamma_-^q
+\frac{1}{6}s\left(\frac{1}{2}g_{pq}R_-
-g_{pi}g_-^{ij}R^-_{jq}
\right)\gamma_+^p\gamma_-^q
\nonumber\\
&&
+\frac{1}{4}tg_{pq}\gamma_-^{q}\gamma_+^{pij}\cR^+_{ij}
+\frac{1}{4}sg_{pq}\gamma_+^{p}\gamma_-^{qij}
\cR^-_{ij}
+S_+S_-
-\frac{1}{2}tg_{pq}\gamma_-^{q}\gamma_+^{p}S_+^2
-\frac{1}{2}sg_{pq}\gamma_+^{p}\gamma_-^{q} S_-^2
\nonumber\\
&&
-\frac{1}{2}tg_{q p}i\gamma_-^{q}\gamma_+^{pj}\nabla_{j }^+ S_+
-\frac{1}{2}sg_{pq}i\gamma_+^{p}\gamma_-^{qj}\nabla^-_jS_-
\nonumber\\
&&
+ts\Biggl[
\frac{1}{12}\left(G^{kl}G^{ij}
+2G^{ik}G^{jl}\right)
\left(R^+_{ij}+R^-_{ij}-2R^g_{ij}\right)
g^+_{p(k}g^-_{l)q}\gamma_+^{p}\gamma_-^q
\nonumber\\
&&
+\frac{1}{8}G^{(ij}G^{kl)}V_{pqijkl}\gamma_+^{p}\gamma_-^q
+\frac{3}{4}N^{jkl}
\left(g^+_{mp}W_+{}^m{}_{(jk}g^-_{l)q}
+g^-_{mq}W_-{}^m{}_{(jk} g^+_{l)p}
\right) 
\gamma_+^{p}\gamma_-^q
\nonumber\\
&&
+\frac{1}{2} M^{kl}g^+_{p(k}g^-_{l)q}\gamma_+^{p}\gamma_-^q
-\frac{3}{4} G^{(ij}G^{kl)}g^+_{jp}g^-_{qi}
[\gamma_+^p,\gamma_-^q] 
\nabla^{g,\cA}_{k}(\cC^+_{l}-\cC^-_{l})
\nonumber\\
&&
%
%
-\frac{3}{4} \Bigl[
G^{(ij}G^{kl)}\left(g^+_{mp}W_+^m{}_{ij}g^-_{kq}
+g^+_{kp}g^-_{mq}W_-^m{}_{ij}
\right)
+N^{jkl}g^+_{pk}g^-_{jq} 
\Bigr]
[\gamma_+^p,\gamma_-^{q}](\cC^+_{l}-\cC^-_{l})
\nonumber\\
&&
+\frac{3}{4} g^+_{pi}g^-_{jq}
G^{(ij}G^{kl)}
\Bigl[\left(\cC^+_{k}\cC^+_{l}+\cC^-_{k}\cC^-_{l}\right)
\left(
\gamma_+^p\gamma_-^q
+\gamma_-^q\gamma_+^p 
\right)
\nonumber\\
&&
-2\cC^+_{k}\cC^-_{l}\gamma_+^p\gamma_-^q
-2\cC^-_{l}\cC^+_{k}\gamma_-^q\gamma_+^p
\Bigr]
\Biggr]
\Biggr\}.
\label{150zax}
\eea

\end{theorem}

\begin{corollary}
\label{corollary3}
The first two coefficients of the asymptotic expansion of the 
relative spectral invariant $\Phi(t,s)$
are
\bea
\Phi_0(t,s) &=& \int\limits_M dx\;\Biggl\{
\frac{n}{2}(t+s)^{-1-n/2}\left(g_+^{1/2}+g^{1/2}_-\right)\tr I
\nonumber\\
&&
-\frac{1}{2}
\left[g^{1/2}(t,s)g_{ij}(t,s)+g^{1/2}(s,t)g_{ij}(s,t)\right]
\tr\,\left(\gamma_+^{i}\gamma_-^{j}\right)
\Biggr\},
\label{141via}\\
\Phi_1(t,s) &=&
\int_M dx\;\Biggl\{
-g^{1/2}(t,s)c_1(t,s)-g^{1/2}(s,t)c_1(s,t)
\label{542saaa}\\
&&
+\left(\frac{n}{2}-1\right)(t+s)^{-n/2}
\Biggl[g_+^{1/2}\tr\left(\frac{1}{6}R_+I-Q_+\right)
+g_-^{1/2}
\tr\Biggl(\frac{1}{6}R_-I-Q_-\Biggr)
\Biggr]\Biggr\},
\nonumber
\eea
\end{corollary}
where $Q_\pm$ are given by (\ref{qsxxab}).

\section{Bogolyubov Invariant}

We motivate the definition of the relative spectral 
invariants by quantum field 
theory. We will be very brief here, the detailed exposition will appear 
elsewhere \cite{avramidi19a}. We 
describe now the standard method for calculation of particles creation via the 
Bogolyubov transformation \cite{birrel80,dewitt75}. 
Let $({\cal M}, h)$ be a pseudo-Riemannian 
$(n+1)$-dimensional 
assume that $({\cal M}, h)$ is globally hyperbolic so that there is a foliation 
of ${\cal M}$ with space slices $M_t$ at a time $t$, moreover, we assume that 
there is a global time coordinate $t$ varying from $-\infty$ to $+\infty$ and 
that at all times $M_t$ is a compact $n$-dimensional Riemannian manifold 
without boundary. We will also assume that there are well defined limits 
$M_{\pm}$ as $ t\to \pm\infty$. For simplicity, we will just assume that the 
manifold ${\cal M}$ has two cylindrical ends, $(-\infty,\beta)\times M$ and 
$(\beta,\infty)\times M$ for some positive parameter $\beta$. So, the foliation 
slices $M_t$ depend on $t$ only on a compact interval $[-\beta,\beta]$. Let 
$\cW$ be a Hermitian vector bundle over $\cM$ and $\cV_t$ be the corresponding 
time slices (vector bundles over $M_t$). 

In quantum field theory there are two types of particles, bosons and fermions. 
The bosonic fields are described by second order Laplace type partial 
differential operators whereas the fermionic fields are described by first 
order Dirac type partial differential operators. Let $L_t$ be a one-parameter 
family of {\it positive self-adjoint elliptic second-order} partial 
differential operators of Laplace type acting on smooth sections of the vector 
bundle $\cV_t$. We assume that there are well defined limits $L_{\pm}$ as $t 
\to \pm\infty$. Let $D_t$ be a  one-parameter family of {\it self-adjoint 
elliptic first-order} partial differential operators of Dirac type acting on 
sections of the vector bundles $\cV_{t}$ such that its square $ L_t=D_t^2 $ is 
a self-adjoint second-order positive elliptic partial differential operator of 
Laplace type. We assume that there are well defined limits $D_{\pm}$ as $t\to 
\pm\infty$. 
Then one defines so-called the in-vacuum and the out-vacuum
and the corresponding in-particles and out-particles. 
Then the out-vacuum contains some in-particles (and vice versa).  
The total number of in-particles in the out-vacuum  is determined by 
the so-called {Bogolyubov invariant}.

Let $E_{b,f,0}$ be the functions defined by
\bea
E_f(x) &=& \frac{1}{e^x+1},
\\
E_b(x) &=& \frac{1}{e^x-1},
\\
E_0(x) &=& \frac{1}{2\sinh x},
\eea
and $\omega_\pm$ are pseudo-differential operators defined by
\be
\omega_\pm=\sqrt{L_\pm}.
\ee
Then in some approximation (for details, see \cite{avramidi19a})
the Bogolyubov invariants for bosons and fermions
are determined by the following traces
\bea
B_b(\beta)&=&\Tr
\left\{E_f(\beta\omega_+)-E_f(\beta\omega_-)\right\}
\Bigl\{E_b(\beta\omega_+)-E_b(\beta\omega_-)\Bigr\},
\label{311xxa}
\\
B_f(\beta) &=& 
2\beta^2\Tr\Bigl\{D_+E_0(\beta\omega_+)-D_-E_0(\beta\omega_-)\Bigr\}^2.
\label{318xxa}
\eea

The Bogolyubov invariants can be expressed in terms of the
the {\it relative spectral invariants}
$\Psi(t,s)$ and $\Phi(t,s)$ defined in
(\ref{220ssb}) and (\ref{248ssb}).
Let $h_{b,f,0}$ be the functions defined by
\bea
h_f(t) &=& 
\frac{1}{2\pi }\fint_\RR dp\; p
\tan\left(\frac{p}{2}\right)\exp(-tp^2)
\nonumber\\
&=&
(4\pi)^{-1/2}t^{-3/2}
\sum_{k=1}^\infty (-1)^{k+1} k\exp\left(-\frac{k^2}{4t}\right),
\\
h_b(t) &=&
\frac{1}{2\pi }\fint_\RR dp\;p
\cot\left(\frac{p}{2}\right)\exp(-tp^2)
\nonumber\\
 &=& (4\pi)^{-1/2}t^{-3/2}
\sum_{k=1}^\infty k\exp\left(-\frac{k^2}{4t}\right),
\\
h_0(t) &=& 
\frac{1}{2\pi }\fint_\RR dp\;
\frac{p}{\sin p}\exp(-tp^2)
\nonumber\\
&=&
(4\pi)^{-1/2}t^{-3/2}
\sum_{k=0}^\infty \left( 2k+1\right)
\exp\left(-\frac{\left(2k+1\right)^2}{4t}\right).
\eea
where the integrals are taken in the principal value sense.
Then the Bogolyubov invariants take the form
\bea
B_b(\beta) &=& 
\int\limits_0^\infty dt\int\limits_0^\infty ds\;
h_f\left(s\right)
h_b\left(t\right)
\Psi\left(\beta^2s,\beta^2t\right),
\label{413xxa}
\\
B_f(\beta) &=&
\int\limits_0^\infty dt\int\limits_0^\infty ds\;\,
h_0\left(s\right)
h_0\left(t\right)
2\beta^2\Phi\left(\beta^2t,\beta^2s\right).
\label{414xxa}
\eea

It is the relative spectral 
invariants $\Psi(t,s)$ and $\Phi(t,s)$ 
that we study in the present paper.
Obviously, the combined heat traces $X(t,s)$ and $Y(t,s)$
contains information about the spectra of both operators $L_\pm$
(and $D_\pm$)
since, in particular,
\bea
X(0,s) &=& \Theta_-(s), \qquad X(t,0)=\Theta_+(t),
\\
Y(0,s) &=& H_-(s), \qquad Y(t,0)=H_+(t)
\eea
Also, although for any $t,s>0$
\bea
\Psi(0,s)=\Psi(t,0)=
\Phi(0,s)=\Phi(t,0)=0,
\eea
the asymptotics as $t,s\to 0$ are non-trivial. It is these asymptotics
that we study in the present paper. 


We can also define the corresponding 
{\it relative zeta functions}
\bea
Z_\Psi(p,q) &=&
\frac{1}{\Gamma(p)\Gamma(q)}\int\limits_0^\infty dt \int\limits_0^\infty 
ds\;t^{p-1}s^{q-1}
\Psi(t,s),
\\
Z_\Phi(p,q) &=&
\frac{1}{\Gamma(p)\Gamma(q)}\int\limits_0^\infty dt \int\limits_0^\infty 
ds\;t^{p-1}s^{q-1}
\Phi(t,s),
\eea
and, similarly, $Z_X(p,q)$ and $Z_Y(p,q)$. Then
\bea
Z_X(p,q) &=& \Tr L_+^{-p}L_-^{-q},
\\
Z_Y(p,q) &=& \Tr D_+^{-2p+1}D_-^{-2q+1}
\eea
and 
\bea
Z_\Psi(p,q) &=& \Tr \left(L_+^{-p}-L_-^{-p}\right)
\left(L_+^{-q}-L_-^{-q}\right),
\\
Z_\Phi(p,q) &=& \Tr  \left(D_+^{-2p+1}-D_-^{-2p+1}\right)
\left(D_+^{-2q+1}-D_-^{-2q+1}\right).
\eea
To avoid confusion the complex
power of the operator $D_\pm$ (which is not positive)
is defined as follows
$D_\pm^{-2p+1}=D_\pm(D_\pm^2)^{-p}$.


For the Dirac case
one can also introduce more general traces
\bea
W_\pm(t,\alpha) &=& \Tr \exp(-t D^2_\pm+i\alpha D_\pm),
\label{424zzc}
\\
V(t,s;\alpha,\beta) &=& \Tr\exp(-tD^2_++i\alpha D_\pm)\exp(-sD^2_-+i\beta D_-).
\label{51xxa}
\eea
Then, obviously,
\bea
\Theta_\pm(t) &=& W_\pm(t,0),
\\
X(t,s) &=& V(t,s;0,0),
\\
Y(t,s) &=& -\frac{\partial}{\partial \alpha}
\frac{\partial}{\partial \beta} V(t,s;\alpha,\beta)
\Big|_{\alpha=\beta=0}.
\eea
Therefore, all traces can be obtained from the traces 
(\ref{424zzc}) and (\ref{51xxa}).

Notice that the trace $W(t,\alpha)$ 
can be written in the form
\be
W_\pm(t,\alpha)=(4\pi t)^{-1/2}\int\limits_\RR d\alpha' 
\exp\left\{-\frac{(\alpha-\alpha')^2}{4t}\right\}
T_\pm(\alpha'),
\label{278gil}
\ee
where
\be
T_\pm(\alpha)=\Tr\exp\left(i\alpha D_\pm\right);
\ee
strictly speaking, $T_\pm(\alpha)$ is a distribution
and eq. (\ref{278gil})
should be understood in the distributional sense.
Similarly, the invariant $V(t,s;\alpha,\beta)$ 
can be written in the form
\be
V(t,s;\alpha,\beta)=(4\pi)^{-1} (ts)^{-1/2}\int\limits_{\RR^2} d\alpha'd\beta' 
\exp\left\{-\frac{(\alpha-\alpha')^2}{4t}
-\frac{(\beta-\beta')^2}{4s}\right\}
S_\pm(\alpha',\beta'),
\ee
where
\be
S_\pm(\alpha,\beta)=\Tr\exp\left(i\alpha D_+\right)
\exp\left(i\beta D_-\right).
\ee

In this paper we will be interested primarily in the asymptotic
expansion of the combined heat traces as $t,s\to 0$.

\section{Generalized Heat Traces}
\label{secxxx}
\setcounter{equation}0

\subsection{Differential Operators}

Let $M$ be a compact $n$-dimensional Riemannian manifold without boundary. 
Throughout the whole paper we denote tensor indices by Latin letters
and use Einstein summation convention. 
We use parenthesis for the symmetrization of indices and square brackets
for the anti-symmetrization. The indices excluded from the symmetrization
or anti-symmetrization are separated by vertical lines.
Also, we denote the
local coordinates by $x^i$ and the partial derivatives by
$\partial_i$.
Let 
$\cV$ be a vector bundle of densities of weight $1/2$ over $M$, $L^2(\cV)$ be 
the corresponding Hilbert space;
we use the notation $\tr$ for the fiber trace and 
$\Tr$ be the corresponding 
$L^2$ trace. We 
study  {\it positive self-adjoint elliptic second-order} partial differential 
operators $L$ with a scalar positive definite leading symbol of {\it Laplace 
type} acting on smooth sections of the bundle $\cV$. A Laplace type operator 
$L$ naturally defines a Riemannian metric $g$ and a connection $\nabla^{\cA}$ 
on the vector bundle with a connection one-form $\cA_i$. Since we will be 
working with different operators we do not have a single metric, then, 
following \cite{avramidi04}, we prefer to work with the vector bundle of 
densities of weight $1/2$ and with the Lebesgue measure $dx$ instead of the 
Riemannian one. Then the heat kernel  $U(t;x,x')$ of the heat semigroup 
$\exp(-tL)$ is also a density of weight $1/2$ at each point $x$ and $x'$, and 
the heat kernel 
diagonal $U(t;x,x)$ is a density of weight $1$. Then a Laplace type 
operator has the form
\bea
L &=& g^{1/4}\left(-\Delta^{g,\cA} +Q\right)g^{-1/4},
\label{lap}
\eea
where $\Delta^{g,\cA}=g^{ij}\nabla^{g,\cA}_i\nabla^{g,\cA}_j$
is the Laplacian,
$g=\det g_{ij}$, 
and $Q$ is some smooth
endomorphism of the vector bundle $\cV$;
locally it has the form
\bea
L &=& -g^{-1/4}(\partial_i+\cA_i)g^{1/2}g^{ij}
(\partial_j+\cA_j)g^{-1/4}+Q.
\eea

Let $L_\pm$ be two {\it Laplace type} operators defined by the metrics 
$g^\pm_{ij}$, the connections $\cA_i^\pm$ and the potential terms $Q_\pm$. 
By using the  metric $g_{ij}(t,s)$, (\ref{113via}), the
connection $\cA_i(t,s)$, (\ref{114via}), and the
identity
\be
tg_+^{ij}\cC_j^+
+sg_-^{ij}\cC_j^-=0
\ee
one can rewrite now the operators $L_\pm$ 
in the form
\bea
L_\pm &=& 
g^{1/4}\left(-\nabla^{g,\cA}_ig_\pm^{ij}\nabla^{g,\cA}_j
-g_\pm^{ij}\cC^\pm_i\nabla^{g,\cA}_j
-\nabla^{g,\cA}_i g_\pm^{ij}\cC^\pm_j
+q_\pm\right)g^{-1/4},
\eea
where
\bea
q_\pm =Q_\pm
-g_\pm^{ij}\cC^\pm_i\cC^\pm_j
+\frac{1}{2}\nabla^g_i(g_\pm^{ij}W^\pm_{j})
+\frac{1}{4}g_\pm^{ij}W^+_{i}W^+_{j},
\eea
where $W^\pm_j$ is defined by (\ref{130via}).

Notice that the sum of Laplace type 
operators is a Laplace type operator, in particular,
the operator
\bea
L(t,s) &=& tL_++sL_-
\nonumber\\
&=&g^{1/4}\left(-\Delta^{g,\cA}+Q\right)g^{-1/4}
\eea
is a Laplace type operator 
with the metric $g_{ij}(t,s)$, (\ref{113via}), the
connection $\cA_i(t,s)$, (\ref{114via}), 
and the potential form 
\bea
Q(t,s)&=&tQ_++sQ_-
-tg_+^{ij}\cC^+_i\cC^+_j
-sg_-^{ij}\cC^-_i\cC^-_j
\label{25zaa}\\
&&
+\frac{1}{2}t\nabla_j^g(g_+^{ij}W^+_{i})
+\frac{1}{4}tg_+^{ij}W^+_{i}W^+_{j}
+\frac{1}{2}s\nabla_j^g(g_-^{ij}W^-_{i})
+\frac{1}{4}sg_-^{ij}W^-_{i}W^-_{j}.
\nonumber
\eea


Now, assume that $\cV$ is a Clifford bundle. Let $D_\pm$ be two {\it 
self-adjoint first-order elliptic} partial differential operators  of {\it 
Dirac 
type} acting on sections of the bundle $\cV$ such that their squares 
$D_\pm^2$ are self-adjoint second-order positive elliptic partial 
differential operators. 
Let $\gamma_\pm: T^*M\to \End(\cV)$ be the Clifford maps 
(determined by traceless Dirac matrices) satisfying 
\be
\gamma_\pm^i\gamma_\pm^j+\gamma_\pm^j\gamma_\pm^i=2 g^{ij}_\pm I,
\label{39viax}
\ee
where $I$ is the identity endomorphism. 
Let $\cA^\pm_i$ be connection one-forms on the vector bundle $\cV$;
it is required to satisfy
\be
\partial_i\gamma_\pm^k+\Gamma_\pm{}^k{}_{ij}\gamma_\pm^j
+[\cA^\pm_i,\gamma_\pm^k]=0,
\label{310viax}
\ee
where $\Gamma_\pm{}^k{}_{ij}$ are Christoffel symbols of the metric $g^\pm_{ij}$,
in particular, it means
$\nabla^\pm_i\gamma_\pm^k=0$.
Let $S_\pm$ be some endomorphisms of the vector bundle $\cV$
anticommuting with $\gamma_\pm^i$.
By using the representation of the
Dirac matrices in terms of the orthonormal frames,
$\gamma_\pm^i(x)=e_\pm{}^i{}_a(x)\gamma^a$, this means that
the matrices $S_\pm$ anti-commute also with the Dirac matrices
$\gamma_\mp^i$, that is,
\be
[S_\pm,\gamma_\pm^i]=[S_\pm,\gamma_\mp^i]=0.
\label{312viax}
\ee
Then the Dirac type operators have the form
\bea
D_\pm &=& g_\pm^{1/4}i\gamma^j_\pm(\partial_j+\cA_j^\pm)g_\pm^{-1/4}+S_\pm
\\
 &=& g_\pm^{1/4}\left(i\gamma^j_\pm\nabla^\pm_j +S_\pm\right)g_\pm^{-1/4}.
\eea
and 
$D_\pm^2$ is a Laplace type operator of the form
\be
D_\pm^2=g_\pm^{1/4}\left(-\Delta_\pm +Q_\pm\right)g_\pm^{-1/4},
\ee 
where
\be
Q_\pm=-\frac{1}{2}\gamma^{ij}_\pm\cR^\pm_{ij}+S_\pm^2+i\gamma_\pm^j\nabla^\pm_j 
S_\pm.
\label{qsxxa}
\ee
$\gamma_\pm^{ij}=\gamma_\pm^{[i}\gamma_\pm^{j]}$
and $\cR^\pm_{ij}$ is the curvature of the connection $\cA^\pm_i$.

If the Clifford bundle is a twisted spinor bundle then
the connection $\cA^\pm_i$ has the form
\be
\cA^\pm_i=\frac{1}{4}\omega^\pm_{abi}\gamma^{ab}+\cE^\pm_i ,
\ee
where
$\omega^\pm_{abi}$ is the spin connection,
and the curvature has the form
\be
\cR^\pm_{ij}=\frac{1}{4}R^\pm_{abij}\gamma^{ab}+\cF^\pm_{ij},
\ee
where
$\cF_{ij}^\pm$ is the curvature of the connection $\cE_{i}^\pm$ and
$R^\pm_{abij}$ is the Riemann tensor of the metric $g^\pm_{ij}$.

\subsection{Heat Traces}

Let $\{\lambda_k^\pm\}_{k=1}^\infty$ be the eigenvalues (counted with 
multiplicities and ordered in nondecreasing order) and 
$\{\varphi_k^\pm\}_{k=1}^\infty$ be the corresponding orthonormal sequence of 
eigensections of the operator $L_\pm$. The heat kernel of the operator $L_\pm$ 
has the following spectral representation
\be
U_\pm(t;x,x')=\sum_{k=1}^\infty \exp\left(-t\lambda^\pm_k\right)
\varphi^\pm_k(x)\varphi^{\pm*}_k(x').
\ee
Then the classical heat trace (\ref{237ssb}) has 
form
\bea
\Theta_{\pm}(t)&=&
\sum_{k=1}^\infty \exp\left(-t\lambda^\pm_k\right)
\label{532xxc}
\nonumber\\
&=&\int\limits_M dx\;\tr U_\pm(t;x,x)
\label{532xxca}
\eea
and the combined heat trace (\ref{238ccx}) is
\bea
X(t,s)&=&
\sum_{k,j=1}^\infty \exp\left(-t\lambda^+_k-s\lambda^-_j\right)
\left|(\varphi^-_j,\varphi^{+}_k)\right|^2
\label{533xxc}
\nonumber\\
&=&
\int\limits_{M\times M}dx\;dx'\;\tr\left\{ U_+(t;x,x')U_-(s;x',x)\right\}.
\label{533xxca}
\eea

Let $\{\mu_k^\pm\}_{k=1}^\infty$ be the eigenvalues of the operator $D_\pm$ 
(counted with multiplicities and ordered in nondecreasing order of the absolute 
value) and $\{\varphi_k^\pm\}_{k=1}^\infty$ be the corresponding orthonormal 
sequence of eigensections of the operator. 
The integral kernel of the heat semigroups
$\exp(-tD^2_\pm+i\alpha D_\pm)$ and $\exp(-tD^2_\pm)$
have the form
\bea
V_\pm(t,\alpha;x,x')
&=&\sum_{k=1}^\infty \exp\left[-t(\mu^\pm_k)^2+i\alpha\mu_k\right]
\varphi^\pm_k(x)\varphi^{\pm*}_k(x'),
\\
U_\pm(t;x,x')&=& \sum_{k=1}^\infty \exp\left[-t(\mu^\pm_k)^2\right]
\varphi^\pm_k(x)\varphi^{\pm*}_k(x').
\eea
Then the classical heat trace (\ref{12via})  has the form
\bea
H_\pm(t) &=& \sum_{k=1}^\infty \mu^\pm_k\exp\left[-t(\mu^\pm_k)^2\right]
\nonumber\\
&=& 
\int\limits_{M}dx\;\tr\left\{ D_\pm U_\pm(t;x,x)\right\},
\eea
where the operators $D_\pm$ act {\it only} on the first argument of the heat kernel,
and the combined heat trace (\ref{18ssb}) is
\bea
Y(t,s)&=& 
\sum_{k,j=1}^\infty \exp\left[-t(\mu^+_k)^2-s(\mu^-_j)^2\right]
\mu^+_k\mu^-_j\left|(\varphi^-_j,\varphi^{+}_k)\right|^2
\nonumber\\
&=&\int\limits_{M\times M}dx\;dx'\;\tr\left\{
 D_+U_+(t;x,x')D_-U_-(s;x',x)\right\},
\label{534xxc}
\eea
where the differential operators act on the first spacial
argument of the heat kernel. 

The generalized traces (\ref{424zzc}) and (\ref{51xxa}) have the form
\bea
W_\pm(t,\alpha)&=& 
\sum_{k=1}^\infty \exp\left[-t(\mu^\pm_k)^2+i\alpha\mu^\pm_k\right],
\nonumber\\
&=&\int\limits_M dx\;\tr V_\pm(t,\alpha;x,x),
\label{56zzaa}
\\
V(t,s;\alpha,\beta)&=&
\sum_{k,j=1}^\infty \exp\left[-t(\mu^+_k)^2+i\alpha\mu^+_k
-s(\mu^-_j)^2+i\beta\mu_j^-\right]
\left|(\varphi^-_j,\varphi^{+}_k)\right|^2.
\nonumber\\
&=&\int\limits_{M\times M}dx\;dx'\tr\left\{
 V_+(t,\alpha;x,x')V_-(s,\beta;x',x)\right\}.
\label{534xxda}
\eea

We would like to stress that whereas the classical invariants $\Theta_\pm(t)$, 
$H_\pm(t)$ and $W_\pm(t)$ depend only on the eigenvalues of the operators the 
new invariants $X(t,s)$, $Y(t,s)$ and $V(t,s;\alpha,\beta)$ depend on the 
eigenfunctions as well and, therefore, contain much more information about the 
spectra of these operators.


\section{ Ruse-Synge  Function}
\setcounter{equation}0

In this section we follow our books \cite{avramidi00,avramidi15}. We fix the 
notation for the rest of the paper. 
Let $x'$ be a {\it fixed point} in a manifold $M$.
We denote indices of tensors in the tangent space at the point $x'$ by prime 
Latin letters. The derivatives with respect to coordinates $x'^i$ will be 
denoted by prime indices as well. 
We will also 
use the notation for the {\it partial derivatives} of a scalar function $f$ with 
respect to $x$ and $x'$ by just adding indices to the function after comma, e.g. 
$f_{,ij'}=\partial_i\partial_{j'}f$.
Obviously, the derivatives 
with respect to $x$ and with respect to $x'$ commute. Finally, everywhere below 
the square brackets denote the diagonal value of a two-point function $f(x,x')$, 
that is, $[f]=f(x',x')$. It is also easy to see that the derivatives of the 
coincidence limits are equal to the sum of the conicidence limits of the 
derivative with respect to $x$ and $x'$
\be
[f]_{,j} =[f_{,j}]+[f_{,j'}].
\label{51qqq}
\ee

Let $g$ be a Riemannian metric
and $r_{\rm inj}(M,g)$ be the injectivity radius of the manifold $M$. 
Let $B_r(x')$ be the geodesic 
ball of radius $r$ less than the injectivity radius of the manifold, $r<r_{\rm 
inj}(M,g)$. Let $U\subset B_r(x')$ be a sufficiently small neighborhood of the 
point $x'$ in the ball $B_r(x')$ so that it is covered by a single coordinate 
patch with coordinates $x^i$. 

Each point $x$ in the neighborhood $U$ can be connected with the
point $x'$ by a {\it unique} geodesic. 
The \index{ Ruse-Synge  function} {\it  Ruse-Synge  function} $\sigma(x,x')$ is
a symmetric smooth function defined as one half of the
square of the geodesic distance $d(x,x')$ between the points $x$ and $x'$,
\be
\sigma(x,x')=\frac{1}{2}d^2(x,x');
\ee
it was introduced by Ruse \cite{ruse31}
and used extensively by Synge  \cite{synge60} and others
\cite{dewitt75,birrel80}
in general relativity under the name {\it world function}.
There are many ways to show that  
the   Ruse-Synge  function satisfies the (modified)
Hamilton-Jacobi equation
\be
\sigma=
\frac{1}{2}g^{ij}(x)\sigma_{,i}\sigma_{,j}
=\frac{1}{2}g^{i'j'}(x')\sigma_{,i'}\sigma_{,j'}
\,,
\label{3133xx}
\ee
with the initial conditions
\be
[\sigma]=
[\sigma_{,i}]
=[\sigma_{,i'}]
=0\,.\qquad
\label{3130zza}
\ee
Furthermore, by differentiating eq. (\ref{3133xx})
and taking the coincidence limit
it is easy to see that
\be
[\sigma_{,ij}] = [\sigma_{,i'j'}] = -[\sigma_{,ij'}] =  g_{ij}.
\label{45via}
\ee

The Hamilton-Jacobi equation (\ref{3133xx}) with  the 
above initial conditions (\ref{3130zza}) has a 
unique solution; it can be solved, for example,
in form of a (noncovariant) Taylor series
\be
\sigma(x,x')=
\sum_{k=2}^\infty\frac{1}{k!}
[\sigma_{,i_1\dots i_k}](x')y^{i_1}\cdots y^{i_k},
\label{36ttt}
\ee
where $y^i=x^i-x'^{i}$.
The coincidence limits of {partial derivatives} of higher orders 
$[\sigma_{,i_1\dots i_k}]$, $k\ge 3$, are {uniquely} determined in terms of 
some 
polynomials in the partial derivatives of the metric $g_{ij, m_1,\dots m_p}$ 
and 
the metrics $g_{ij}$ and $g^{ij}$, that is, some polynomials in the partial 
derivatives $[\sigma_{,ij}]_{, m_1,\dots m_p}$ and the matrix $[\sigma_{,ij}]$ 
and its inverse. Therefore, there are { non-trivial relations} between the 
coincidence limits of partial derivatives.
By using these equations one can obtain the coincidence limits 
of partial derivatives
\bea
[\sigma_{,ijk}] &=& 
3 g_{m(k}\Gamma{}^m{}_{ij)}=
\frac{3}{2}g_{(ij,k)},
\\
{}[\sigma_{,ijkl}] &=&
4g_{m(l}\Gamma^m{}_{ij,k)} 
+4g_{m(l}\Gamma^n{}_{ij}\Gamma^{m}{}_{k)n}
+3g_{nm}\Gamma^n{}_{(ij}\Gamma^{m}{}_{kl)},
\label{48viax}
\\
{}[\sigma_{,i'jkl}] &=&
-g_{m(l}\Gamma^m{}_{ij,k)} 
-g_{m(l}\Gamma^n{}_{ij}\Gamma^{m}{}_{k)n},
\label{49viax}
\eea
where $\Gamma^i{}_{jk}$ are the Christoffel symbols for the metric $g$. Here 
and 
everywhere below the parenthesis denote the symmetrization over all included 
indices and the vertical lines denote the indices excluded from the 
symmetrization.

By differentiating eq. (\ref{3133xx}) we also find
\be
\sigma_{,k'}=g^{ij}\sigma_{,ik'}\sigma_{,j}.
\ee 
Let $\gamma^{j'i}$ be the inverse of the matrix of mixed
derivatives $\sigma_{,jk'}$
(it should not be confused with Dirac matrices).
Then we obtain
\be
\gamma^{k'i}\sigma_{,k'}=g^{ij}\sigma_{,j},
\label{524zzz}
\ee
and, therefore, the Ruse-Synge function satisfies a
{\it non-trivial equation
without any metric}
\be
\sigma=\frac{1}{2}\gamma^{i'j}\sigma_{,i'}\sigma_{,j}.
\label{525zzz}
\ee


This enables one to compute the Ruse-Synge function in terms of diagonal values 
of its own partial derivatives. The usual Taylor series (\ref{36ttt}) is not 
symmetric whereas the function $\sigma(x,x')$ is. Thus, it is more appropriate 
to represent it in the manifestly symmetric Taylor series. Let us introduce new 
coordinates
\be
z^i=x^i+x'^i, \qquad 
y^i=x^i-x'^i.
\ee
Then the Ruse-Synge function is
a function of $z$ and $y$
\be
\sigma(x,x')=f(z,y).
\ee
Then the derivatives are related by
\bea
\partial^z_i &=& \frac{1}{2}\left(\partial^x_i+\partial^{x'}_i\right),
\qquad
\partial^{y}_i = \frac{1}{2}\left(\partial^x_i-\partial^{x'}_i\right),
\\
\partial^x_i &=& \partial^z_i+\partial^{y}_i,
\qquad
\partial^{x'}_i = \partial^z_i-\partial^{y}_i.
\eea 
We can expand the Ruse-Synge function
in the Taylor series in the variables $y$ with
coefficients depending on the variables $z$.
Since it is symmetric it will only have even powers
of $y$, 
\be
\sigma(x,x')=\sum_{k=1}^\infty \frac{1}{(2k)!}F_{i_1\dots 
i_{2k}}(z)y^{i_1}\dots y^{i_{2k}},
\ee
Then the derivatives of the Ruse-Synge function are
\bea
\sigma_{,i} &=& A_i+B_i,
\\
\sigma_{,j'} &=& A_j-B_j,
\\
\sigma_{,ij'} &=& -F_{ij}+C_{ij}+D_{ij},
\eea
where
\bea
A_j&=& 
\sum_{k=1}^\infty
\frac{1}{(2k)!}F_{i_1\dots i_{2k},j}
y^{i_1}\dots y^{i_{2k}},
\\
B_j &=& \sum_{k=0}^\infty
\frac{1}{(2k+1)!}F_{i_1\dots i_{2k+1}j}y^{i_1}\dots y^{i_{2k+1}},
\\
C_{ij} &=& \sum_{k=1}^\infty
\frac{1}{(2k)!}\left(F_{i_1\dots i_{2k},ij}
-F_{i_1\dots i_{2k}ij}\right)y^{i_1}\dots y^{i_{2k}},
\\
D_{ij}&=&\sum_{k=0}^\infty
\frac{1}{(2k+1)!}\left(F_{i_1\dots i_{2k+1}i,j}
-F_{i_1\dots i_{2k+1}j,i}
\right)y^{i_1}\dots y^{i_{2k+1}}.
\eea
Now, by using these expansions one can compute the expansion of the matrix 
$\gamma^{ij'}$ and then use the equation (\ref{525zzz}) to obtain recursive 
relation for the coefficients $F_{i_1\dots i_k}$. {\it All of the higher-order 
coefficients $F_{i_1\dots i_k}$, with $k\ge 4$, will be determined by the 
derivatives of the first coefficient $F_{ij}$}.

The diagonal values of the {\it covariant derivatives} of the Ruse-Synge 
function are expressed in terms of the polynomials of the covariant derivatives 
of the curvature tensor, in particular, 
\bea
[\nabla^g_{i}\nabla^g_j\nabla^g_{k}\sigma]
&=& [\nabla^g_{i}\nabla^g_j\nabla^g_{k'}\sigma]=0,
\label{324mmm}
\\
{}[\nabla^g_{l}\nabla^g_k\nabla^g_{j}\nabla^g_{i}\sigma]
&=&-[\nabla^g_{l'}\nabla^g_k\nabla^g_{j}\nabla^g_{i}\sigma]
= [\nabla^g_{l'}\nabla^g_{k'}\nabla^g_{j}\nabla^g_{i}\sigma]
=-\frac{2}{3}R^g{}_{(i|k|j)l}.
\label{325mmm}
\eea
That is, {\it the diagonal values of all higher order
covariant derivatives of the Ruse-Synge function 
$[\nabla^g_{j'_m}\cdots\nabla^g_{j'_1}\nabla^g_{i_k}\cdots\nabla^g_{i_1}
\sigma]$, with $k+m\ge 4$,
are expressed in terms of the 
derivatives $[\sigma_{,ij'}]_{,i_1\dots i_k}$ of the
diagonal values of the second derivatives
$[\sigma_{,ij'}]$}.
One can also show that
it also satisfies the following coincidence limits \cite{avramidi00}:
for any $k\ge 2$,
\be
[\nabla^g_{(i_1}\cdots\nabla^g_{i_k)}\nabla^g_{j'}\sigma]=
[\nabla^g_{(i_1}\cdots\nabla^g_{i_k)}\nabla^g_j\sigma]=0.
\ee

An important ingredient is the Van Vleck-Morette determinant, defined by
\bea
M(x,x') &=& \det\left(-\sigma_{,ij'}(x,x')\right)\,;
\label{vvmd}
\eea
it is a  two-point density of weight $1$ at each point
(we denote it by $M(x,x')$ instead of the usual $D(x,x')$
to avoid confusion with the Dirac type operators $D_\pm$).
Therefore, we find it convenient to define the function
\be
\zeta(x,x')
=\frac{1}{2}\log\left(g^{-1/2}(x)M(x,x')g^{-1/2}(x')\right),
\label{410qqq}
\ee 
which is a scalar function at each point.
The first coincidence limits of this functions are
\cite{avramidi00}
\bea
[\zeta] &=& [\zeta_{,i}]=0,
\label{511mmm}
\\
{}[\nabla^g_{i}\nabla^g_{j}\zeta] &=& \frac{1}{6}R^g_{ij},
\label{512mmm}
\\
{}[\nabla^g_{(i}\nabla^g_{j}\nabla^g_{k)}\zeta] &=& \frac{1}{4}\nabla^g_{(i}R^g{}_{jk)},
\label{513via}
\\
{}[\nabla^g_{(i}\nabla^g_{j}\nabla^g_{k}\nabla^g_{l)}\zeta] &=& 
\frac{3}{10}\nabla^g_{(i}\nabla^g_{j}R^g{}_{kl)}
+\frac{1}{15}R^g{}_{m(i}{}^n{}_{j}R^{g}{}_k{}^{m}{}_{l)n}.
\label{514via}
\eea
 

One can also show that the tangent vector to the geodesic 
connecting the points $x'$ and $x$ at the point $x'$ pointing to the
point $x$ is given 
by the derivative of the   Ruse-Synge function \cite{avramidi00}
\be
\xi^{i'}=-g^{i'j'}\sigma_{,j'},
\label{514xxz}
\ee
so that
\be
\sigma=\frac{1}{2}g_{i'j'}\xi^{i'}\xi^{j'}.
\ee 
The variables $\xi^{i'}$ are related to the so called Morse variables; they 
provide the normal coordinates in geometry. The Jacobian of the transformation 
$x\mapsto \xi$ is expressed in terms of the Van Vleck-Morette determinant and 
for sufficiently close points $x$ and $x'$ is not equal to zero. The volume 
element and the derivatives in these coordinates have the form
\bea
dx &=& M^{-1}(x,x')g(x')d\xi
\nonumber\\
&=& g^{1/2}(x')g^{-1/2}(x)e^{-2\zeta(x,x')}\;d\xi.
\label{59qqq}
\\
\frac{\partial}{\partial x^i} &=&
-\sigma_{,ik'}g^{k'j'}\frac{\partial}{\partial \xi^{j'}}\,.
\eea
Then an arbitrary analytic scalar function $f$ can be expanded
in the covariant Taylor series, \cite{avramidi00}
\be
f=\sum_{k=0}^\infty\frac{1}{k!}
f_{i'_1 \dots i'_k}
\xi^{i'_1}\cdots
\xi^{i'_k}\,,
\ee
where $f_{i'_1 \dots i'_k}=
[\nabla^g_{(i_1}\cdots\nabla^g_{i_k)}f](x')$.


One can show that the metric is determined by
the Ruse-Synge function as follows.
Let $V$ be the matrix defined by
\be
V_{k'l'}=\sigma_{,j'}\gamma^{ij'}\sigma_{,k'l'i}.
\ee
Further, let $Y$ be a matrix defined by 
\be
Y_{k'l'}=\sigma_{,k'l'}-V_{k'l'},
\ee
and $X=(X^{i'j'})$ be the inverse of the matrix $Y$. 
Then the matrix $X$ is given by the series
\be
X=\sum_{n=0}^\infty (\beta V)^n\beta,
\ee
where $\beta^{k'l'}$ is the inverse of the matrix $\sigma_{,k'l'}$,
that is,
\bea
X^{k'l'} &=& 
\beta^{k'l'}+
\beta^{k'm'}V_{m'p'}\beta^{p'l'}
+\beta^{k'm'}V_{m'p'}\beta^{p'q'}V_{q'r'}\beta^{r'l'}
+\cdots
\nonumber
\eea

By differentiating eq. (\ref{524zzz}) with respect to $x'^l$
we obtain
\bea
g^{ij}\sigma_{,jl'} 
&=& 
\gamma^{ik'}Y_{k'l'}.
\eea
Finally, by multiplying by the matrix $\gamma^{jl'}$ we
prove the following lemma.

\begin{lemma}
\label{lemmaruse}
The metric is uniquely determined by the partial 
derivatives of the Ruse-Synge function by
\bea
g^{ij} &=&\gamma^{ik'}\gamma^{jl'}Y_{k'l'},
\label{57zzz}
\\
g_{ij} &=& \sigma_{,ik'}\sigma_{,jl'}X^{k'l'}.
\eea
\end{lemma}
\noindent

Even though the metric is determined by the off-diagonal
derivatives of $\sigma$ it does not depend on the point $x'$.
Also, of course for $x=x'$ we get $g_{ij}(x')=g_{i'j'}$.
Notice that the matrix $V$ is of first order in $y^i=x^i-x'^i$;
therefore, this power series is well defined near diagonal.
Thus, we obtain for the metric
\bea
g_{ij} &=& 
\sigma_{,ik'}\sigma_{,jl'}\beta^{k'l'}
+\sigma_{,ik'}\sigma_{,jl'}\beta^{k'm'}V_{m'p'}\beta^{p'l'}
\nonumber\\
&&+\sigma_{,ik'}\sigma_{,jl'}\beta^{k'm'}V_{m'p'}\beta^{p'q'}V_{q'r'}\beta^{r'l'
}
+\cdots
\eea
Therefore, one can find the metric in terms of the Taylor series
\bea
g^{ij}(x) &=& \sum_{k=0}^\infty\frac{1}{k!}
g^{ij}{}_{,i_1\dots i_k}(x')y^{i_1}\cdots y^{i_k},
\\
g_{ij}(x) &=& \sum_{k=0}^\infty\frac{1}{k!}
g_{ij}{}_{,i_1\dots i_k}(x')y^{i_1}\cdots y^{i_k}.
\eea
The Taylor coefficients $g^{ij}{}_{,i_1\dots i_k}(x')$
and $g_{ij}{}_{,i_1\dots i_k}(x')$  
are expressed in terms of polynomials in the
coincidence limits $[\sigma_{,k'i_1\dots i_p}]$
and $[\sigma_{,k'l'i_1\dots i_p}]$ and the metric $g^{i'j'}(x')$.
This gives an expression for the metric entirely in terms
of the partial derivatives of the Ruse-Synge function.


Finally, we study the dependence of the Ruse-Synge function on the metric. Let 
$h_{ij}$ be another metric and $\sigma^h(x,x')$ be the Ruse-Synge function for 
the metric $h$. We will need to study the covariant Taylor expansion of this 
function in the power series in the variables $\xi^{i'}$ defined by 
(\ref{514xxz}) (with respect to the metric $g$), that is,
\be
\sigma^h(x,x')=\sum_{k=2}^\infty \frac{1}{k!}
[\nabla^g_{(i_1}\cdots\nabla^g_{i_k)}\sigma^h](x')\xi^{i'_1}\cdots \xi^{i'_k}.
\ee
To avoid confusion we use the notation $\nabla^g$ and $\nabla^h$ to 
denote the covariant derivatives with respect to the metrics $g$ and $h$. All 
indices will be raised and lowered by the metric $g$.

The function $\sigma^h$ satisfies the equation
\be
\sigma^h=\frac{1}{2}h^{ij}\sigma^h_{,i}\sigma^h_{,j}
\ee
with the initial conditions
\be
[\sigma^h]=[\sigma^h_{,i}]=0;
\ee
therefore, the first two terms in the Taylor series of the function
$\sigma^h$ vanish. 

The non-compatibility of the metric $h$ and $g$ is measured by 
the non-metricity tensor
\be
K_{ijk}=\nabla^g_i h_{jk}
\ee
and the disformation tensor
\be
W^i{}_{jk}=\Gamma_h{}^i{}_{jk}-\Gamma_g{}^i{}_{jk},
\ee
where $\Gamma_{h,g}{}^i{}_{jk}$ are the Levi-Civita connections
of the metrics $h$ and $g$. These two tensors are related by
\bea
W^i{}_{jk} &=& \frac{1}{2}h^{im}\left(K_{jkm}+K_{kjm}-K_{mjk}\right),
\\
K_{ijk} &=& h_{km}W^{m}{}_{ij}  + h_{jm} W^{m}{}_{ik}.
\eea
The covariant derivatives with respect to the metrics $g$ and $h$
are related by
\bea
\nabla^h_i T^k{}_j &=& \nabla^g_i T^k{}_j
+W^k{}_{im}T^m{}_j
-W^m{}_{ij}T^k{}_m.
\label{360mmm}
\eea
We introduce the following scalar 
$W=\frac{1}{2}\log \left(\frac{h}{g}\right)$
with $h=\det h_{ij}$, $g=\det g_{ij}$,
and the vector
$
W_j =\partial_j W,
$
then
\be
W^i{}_{ij}=\frac{1}{2}h^{kl}K_{jkl}=W_j.
\ee
The Riemann tensors  are related by
\bea
R_{h}{}^i{}_{jkl} &=&
R_{g}{}^i{}_{jkl}
+\nabla_k^g W^i{}_{lj}
-\nabla_l^g W^i{}_{kj}
+W^i{}_{km}W^m{}_{lj}
-W^i{}_{lm}W^m{}_{kj}.
\eea
One has to be careful with this equation when lowering or raising indices.
For example, the Ricci tensors are obtained by just contracting the indices
\bea
R^h{}_{jl} &=&
R^g{}_{jl}
+\nabla_i^g W^i{}_{lj}
-\nabla^g_{j}W_{l}
+W_{m}W^m{}_{lj}
-W^i{}_{lm}W^m{}_{ij},
\label{355zaa}
\eea
but for the Riemann tensor $R^h{}_{ijkl}$ with all indices lowered
we have to use the metric $h_{ij}$ and, therefore,
it will not be directly related to the Riemann tensor $R^g{}_{ijkl}$,
which is obtained by using the metric $g_{ij}$, that is,
\be
R^{h}{}_{njkl} =
h_{ni}g^{im}R^{g}{}_{mjkl}
+h_{ni}\nabla_k^g W^i{}_{lj}
-h_{ni}\nabla_l^g W^i{}_{kj}
+h_{ni}W^i{}_{km}W^m{}_{lj}
-h_{ni}W^i{}_{lm}W^m{}_{kj}.
\ee

We will need to compute the following tensors
determined by the diagonal values
of the 
symmetrized covariant derivatives with respect to the metric $g$
of the Ruse-Synge function $\sigma^h$
of the metric $h$ and the vectors $\sigma^h_{,j}$ and $\sigma^h_{,j'}$,
\bea
S_{i_1\dots i_k} &=& [\nabla^g_{(i_1}\cdots\nabla^g_{i_k)}\sigma^h],
\label{449viax}
\\
T_{ji_1\dots i_k} &=& [\nabla^g_{(i_1}\cdots\nabla^g_{i_k)}\nabla^g_j\sigma^h],
\\
V_{ji_1\dots i_k} &=& [\nabla^g_{(i_1}\cdots\nabla^g_{i_k)}\nabla^g_{j'}\sigma^h].
\eea
We know that
\bea
[\nabla^h_{i}\sigma^h] &=& 0,
\\
{}[\nabla^h_{j}\nabla^h_{i}\sigma^h] &=& h_{ij},
\label{461viax}
\\
{}[\nabla^h_{k}\nabla^h_{j}\nabla^h_{i}\sigma^h] &=& 
[\nabla^h_{k}\nabla^h_{j}\nabla^h_{i'}\sigma^h] 
= 0,
\label{462viax}\\
{}[\nabla^h_{l}\nabla^h_{k}\nabla^h_{j}\nabla^h_{i}\sigma^h] 
&=& -{}[\nabla^h_{l'}\nabla^h_{k}\nabla^h_{j}\nabla^h_{i}\sigma^h] 
=-\frac{2}{3}R^h{}_i{}_{(k|j|l)}.
\eea
By using these equations and the relation (\ref{360mmm})
between the covariant derivatives we obtain
\bea
[\nabla^g_i \sigma^h] &=& S_i=T_i=V_i=0,
\\
{}[\nabla^g_j\nabla^g_i\sigma^h] &=& -[\nabla^g_j\nabla^g_{i'}\sigma^h] = 
S_{ij}=T_{ij}=-V_{ij}=
h_{ij},
\label{465viax}
\\{}
{}[\nabla^g_k\nabla^g_j\nabla^g_i \sigma^h] &=& 
S_{ijk}=T_{ijk}
=3h_{m(i}W^m{}_{jk)}
=\frac{3}{2}K_{(ijk)}.
\label{361zaa}
\eea
Also, by using the relation (\ref{51qqq}) and (\ref{361zaa})
(or (\ref{462viax}) and (\ref{360mmm})) we obtain
\bea
[\nabla^g_k\nabla^g_j\nabla^g_{i'} \sigma^h] &=& 
V_{ijk}
=-h_{mi}W^m{}_{kj}.
\label{467viax}
\eea
Similarly, we compute
\bea
[\nabla^g_{(l}\nabla^g_k\nabla^g_{j)}\nabla^g_{i} \sigma^h]
&=& T_{ijkl}
\nonumber\\
&=&
3h_{m(j} \nabla^g{}_{k}W^m{}_{l)i}
+h_{mi} \nabla^g{}_{(j}W^m{}_{kl)}
+3h_{m(j}W^n{}_{k|i|}W^{m}{}_{l)n}
\nonumber\\
&&
+h_{mi}W^{n}{}_{(jk}W^{m}{}_{l)n}
+
3h_{nm}W^n{}_{(jk}W^{m}{}_{l)i},
\label{469viax}\\
{}[\nabla^g_{(l}\nabla^g_k\nabla^g_{j)}\nabla^g_{i'} \sigma^h]
&=& 
V_{ijkl}
\nonumber\\
&=& -h_{mi} \nabla^g{}_{(j}W^m{}_{kl)}
-h_{mi}W^n{}_{(jk}W^{m}{}_{l)n}.
\label{470viax}
\\
{}[\nabla^g_{(l}\nabla^g_k\nabla^g_j\nabla^g_{i)} \sigma^h]
&=& 
S_{ijkl}
\label{362zaa}\\
&=& 4h_{m(i} \nabla^g{}_{j}W^m{}_{kl)}
+4h_{m(i}W^n{}_{jk}W^{m}{}_{l)n}
+3h_{nm}W^n{}_{(ij}W^{m}{}_{kl)}.
\nonumber
\eea
This can also be written in terms of the tensors $K_{ijk}$
\be
S_{ijkl}
=2\nabla^g{}_{(i}K_{jkl)}
-h^{mn}K_{(ij|m|}K_{kl)n}
+h^{mn}K_{n(ij}K_{kl)m}
-\frac{1}{4}h^{mn}K_{m(ij}K_{|n|kl)}.
\label{362zaaz}
\ee


Let $\nabla^\cA$ be a connection on a vector bundle $\cV$ over a manifold $M$.
It defines the {\it operator of parallel transport} $\cP_{g,\cA}(x,x')$ 
of sections of the vector bundle
$\cV$ along geodesics of the metric $g$
from the point $x'$ to the point $x$.
It satisfies the equation of parallel transport
\cite{avramidi00}
\be
g^{ij}\sigma^g_{,j}\nabla^\cA_i\cP_{g,\cA}=0
\ee
with the initial condition
\bea
[\cP_\cA]&=&I,
\eea
where $I$ is the identity endomorphism.
By using these equations we obtain the coincidence limits of partial
derivatives
\bea
[\cP^{g,\cA}_{,i}] &=& -\cA_i,
\\{}
[\cP^{g,\cA}_{,ij}] &=& 
 -\cA_{(i,j)}+\cA_{(i}\cA_{j)}.
\eea
The coincidence limits of covariant derivatives are
\bea
[\nabla^\cA_i \cP_{g,\cA}] &=&0,
\label{420baa}
\\{}
[\nabla^{g,\cA}_i\nabla^{g,\cA}_j\cP_{g,\cA}] &=& \frac{1}{2}\cR^\cA_{ij},
\label{422qqc}
\eea
where $\cR^\cA_{ij}$ is the curvature of the connection $\nabla^\cA$.
Moreover, one can show that the diagonal values of the 
symmetrized covariant derivatives vanish,
\be
[\nabla^{g,\cA}_{(i_1}\cdots\nabla^{g,\cA}_{i_k)}\cP_{g,\cA}]=0.
\ee

Now, suppose that there is another metric $h$ and another
connection $\nabla^{h,\cB}$ and $\cP_{h,\cB}$
be the corresponding operator of parallel transport. We need to compute the
diagonal values of the 
derivatives $[\nabla^{g,\cA}_{(i_1}\cdots\nabla^{g,\cA}_{i_k)}\cP_{h,\cB}]$.
The difference of the connection one-forms defines the tensor
\be
\cC_i=\cB_i-\cA_i,
\ee
so that
\be
\nabla^{g,\cA}_i\cP_{h,\cB}=\nabla^{h,\cB}_i\cP_{h,\cB}-\cC_i\cP_{h,\cB}.
\label{483viax}
\ee
By using the eqs. (\ref{420baa}), (\ref{422qqc}), (\ref{483viax}) we obtain
\bea
[\nabla^{g,\cA}_{i}\cP_{h,\cB}] &=& -\cC_i,
\label{372zaa}
\\
{}[\nabla^{g,\cA}_{(i}\nabla^{g,\cA}_{j)}\cP_{h,\cB}] &=& 
-\nabla^{g,\cA}_{(i}\cC_{j)}+\cC_{(i}\cC_{j)}.
\label{373zaa}
\eea


\section{Asymptotics of Integrals}
\setcounter{equation}0

We use the Laplace method to compute the asymptotics as $\varepsilon\to 0$
of Laplace type integrals
\be
F(\varepsilon)=
(4\pi\varepsilon)^{-n/2}
\int\limits_{U}dx\;\exp\left(-\frac{1}{2\varepsilon}\Sigma(x,x')\right)
\varphi(x),
\label{541zzam}
\ee
with some positive smooth function $\Sigma$ and a smooth function $\varphi$
over a sufficiently small neighborhood $U$ of a point $x'$
in a manifold $M$. 

\subsection{Gaussian Integrals on Riemannian Manifolds}

First of all, we recall the standard Gaussian integrals.
Let $G_{ij}$ be a real symmetric positive matrix,
$G^{ij}$ be its inverse, $G=\det G_{ij}$ and 
$\left<y, Gy\right>=G_{ij}y^iy^j$. 
We define the {\it Gaussian average} of a 
smooth function $f$ on $\RR^n$ by
\be
\left<f\right>_G =
(4\pi)^{-n/2}G^{1/2}\int\limits_{\RR^n}dy\;
\exp\left(-\frac{1}{4}\left<y, Gy\right>\right)
f(y).
\ee
Then the Gaussian average  of the odd monomials vanish
and the average of the monomials for any $k\ge 0$ are
(see, e.g. \cite{avramidi15, prudnikov83})
\be
\left<y^{i_1}\cdots y^{i_{2k}}\right>_G
= \frac{(2k)!}{k!}
G^{(i_1i_2}\cdots G^{i_{2k-1}i_{2k})}.
\label{53via}
\ee
By integrating by parts it is easy to obtain a useful relation
for the averages of derivatives
\be
\left<\partial_{i_1}\cdots\partial_{i_k}f\right>_G=
\left<\cH_{i_1\dots i_k}f\right>_G,
\ee
where $\cH_{i_1\dots i_k}$ are Hermite polynomials defined by
\cite{erdelyi53}
\bea
\cH_{i_1\dots i_k}(y) &=& (-1)^k
\exp\left(\frac{1}{4}\left<y, Gy\right>\right)
\partial_{i_1}\cdots\partial_{i_k}
\exp\left(-\frac{1}{4}\left<y, Gy\right>\right)
\nonumber\\
&=&
(-1)^k\cD_{i_1}\cdots\cD_{i_k}\cdot 1,
\eea
and
\be
\cD_{i}=\partial_i-\frac{1}{2}G_{ij}y^j.
\ee
As a result, the Gaussian average of a Hermite polynomial
of degree $k$ with any polynomial $f$ of degree less than $k$ vanishes
\be
\left<\cH_{i_1\dots i_k}f\right>_G=0,
\ee
and the average of the product of Hermite polynomials of the same degree
is
\be
\left<\cH_{i_1\dots i_k}\cH_{j_1\dots j_k}\right>_G
=\frac{k!}{2^k}G_{i_1(j_1}\cdots G_{|i_k|j_k)}.
\ee

By the same trick one could get the relations
\bea
\left<y^if\right>_G &=& 2G^{ij}\left<\partial_j f\right>_G,
\\
\left<y^iy^jf\right>_G &=& 2G^{ij}\left<f\right>_G
+4G^{ik}G^{jm}\left<\partial_k\partial_m f\right>_G,
\eea
etc.

\begin{lemma}
\label{lemma1}
Let $U$ be an open set in $\RR^n$ containing the origin and
$\varphi$ be a smooth real function on $U$.
Let $\varepsilon>0$ be a positive real parameter, 
and
\be
F(\varepsilon)
=(4\pi\varepsilon)^{-n/2}
\int\limits_{U}dy\;
\exp\left(-\frac{1}{4\varepsilon}\left<y, Gy\right>\right)\varphi(y).
\ee
Then there is the asymptotic expansion of the integral 
$F(\varepsilon)$
as $\varepsilon\to 0^+$, independent of $U$,
\be
F(\varepsilon)
\sim 
\sum_{k=0}^\infty \varepsilon^{k}c_k,
\ee  
where
\be
c_k =
\frac{1}{k!}
G^{(i_1i_2}\cdots G^{i_{2k-1}i_{2k})}
G^{-1/2}\varphi_{i_1 \dots i_{2k}},
\label{55via}
\ee
and $\varphi_{i_1\dots i_k}=\varphi_{,i_1\dots i_k}(0)$.
\end{lemma}

{\it Remark.} This can also be written as
\be
c_k=\frac{1}{k!}(\Delta_G^k G^{-1/2}\varphi)(0),
\ee
where 
\be
\Delta_{G} = G^{ij}\frac{\partial}{\partial y^i}
\frac{\partial}{\partial y^j}.
\ee

{\it Proof.}
This lemma can be proved by using the Taylor expansion. The open set $U$ must 
contain an open ball $B_\delta(0)$ of some radius $\delta>0$ centered at the 
origin. After rescaling of the variables $y^i\mapsto \sqrt{\varepsilon}\,y^i$ 
the domain of the integration becomes $U_\varepsilon$ containing the ball 
$B_{\delta/\sqrt{\varepsilon}}(0)$ of radius $\delta/\sqrt{\varepsilon}$ and as 
$\varepsilon\to 0$ it becomes the whole space $\RR^n$. The calculation of 
Gaussian average gives then the result.


We will need the following Lemma to compute the coefficients of the asymptotic 
expansion. We use the notation introduced at the beginning of this section.
We pick a metric $g$ and let $R^g{}_{ijkl}$ be the Riemann tensor,
$R^g_{ij}$ be the Ricci tensor, $R_g$ be the 
scalar curvature, $\nabla^g_i$ be the covariant derivative 
(also denoted by the semicolon $;$)
and $\Delta_g$ be the scalar Laplacian of the metric $g$.
Further, let
$\sigma$ be the Ruse-Synge function of this metric and $\zeta$ be the modified Van 
Vleck-Morette determinant.
We generalize the Gaussian integrals in the Euclidean space to Riemannian
manifolds by replacing the quadratic form in the exponential
by the Ruse-Synge function.


\begin{lemma}
\label{lemma5}
Let $U$ be a sufficiently small neighborhood of a fixed point $x'$
in a manifold $M$, $\varphi(x,x')$ be a smooth scalar density of weight $1$, 
and  
\be
F(\varepsilon) 
=(4\pi\varepsilon)^{-n/2}
\int\limits_{U}dx\,
\exp\left(-\frac{\sigma(x,x')}{2\varepsilon}\right)
\varphi(x)\,.
\label{58via}
\ee
Then as $\varepsilon\to 0^+$ there is the asymptotic expansion
independent of $U$
\be
F(\varepsilon)
\sim 
\sum_{k=0}^\infty \varepsilon^{k} c_k,
\ee
where
\be
c_k=\frac{1}{k!}
g^{i_1i_2}\cdots g^{i_{2k-1}i_{2k}}
\hat\varphi_{i_1\dots i_{2k}},
\ee
and 
\be
\hat\varphi_{i'_1\dots i'_k}=\left[
\left(\nabla^g_{(i_1}-2\zeta_{,(i_1}\right)\cdots
\left(\nabla^g_{i_{k})}-2\zeta_{,i_{k})}\right)
(g^{-1/2}\varphi)\right](x').
\ee
The coefficients $c_k$ are polynomial in the derivatives of the curvature
of the metric $g$ and linear in the derivatives of the function $\varphi$.
In particular,
\bea
c_0 &=& g^{-1/2}[\varphi],
\\
c_1 &=& [\Delta_gg^{-1/2}\varphi] -\frac{1}{3}R_gg^{-1/2}[\varphi],
\label{555zaz}
\\
c_2 &=& 
\frac{1}{2}g^{ij}g^{kl}[\nabla^g_{(i}\nabla^g_{j}\nabla^g_{k}\nabla^g_{l)}(g^{-1/2}\varphi)]
-\frac{2}{3}R_g^{ij}[\nabla^g_{(i}\nabla^g_{j)}(g^{-1/2}\varphi)]
\nonumber\\
&&
-\frac{1}{3}R_g[\Delta_g(g^{-1/2}\varphi)]
-\frac{2}{3}R_g{}^{;i}[\nabla^g_{i}(g^{-1/2}\varphi)]
\nonumber\\
&&
+\left(\frac{1}{18}R_g^2
-\frac{1}{5}(\Delta_g R_g)
+\frac{4}{45}R_g{}^{ij}R^{g}{}_{ij}
-\frac{1}{30}R_g{}^{ijkl}R^g{}_{ijkl}\right)g^{-1/2}[\varphi].
\label{522viax}
\eea

\end{lemma}

\noindent
{\it Proof.}
Let $\xi^{i'}=-g^{i'j'}\sigma_{,j'}$ and $|\xi|^2=g_{i'j'}\xi^{i'}\xi^{j'}$.
Then by changing the variables $x\mapsto \xi$ and
using eq. (\ref{59qqq}) we obtain
\bea
F(\varepsilon)
&=&  (4\pi\varepsilon)^{-n/2}
\int\limits_{\hat U} d\xi\,g^{1/2}(x')
\exp\left(-\frac{1}{4\varepsilon}|\xi|^2\right)
\hat\varphi(\xi)
\,.
\eea
where $\hat U$ is the corresponding domain in the variables $\xi$
and
\be
\hat\varphi(\xi)=\exp\left\{-2\zeta(x,x')\right\}g^{-1/2}(x)\varphi(x)\,.
\label{557ccx}
\ee

We rescale the variables $\xi\mapsto\sqrt{\varepsilon}\xi$. Then
as $\varepsilon\to 0$ we can extend the integration
domain to the whole space $\RR^n$; this does not affect 
the asymptotic expansion. 
Therefore, the asymptotic expansion
is determined by the Gaussian average
\be
F(\varepsilon)\sim 
\left<\hat\varphi(\sqrt{\varepsilon}\xi)\right>_{g}.
\ee

Next, we expand the function $\hat\varphi$ 
in the covariant Taylor series
\bea
\hat\varphi(\sqrt{\varepsilon}\xi)
&=&\sum_{k=0}^\infty
\frac{\varepsilon^{k/2}}{k!}
\hat\varphi_{i'_1\dots i'_k}
\xi^{i'_1}\cdots\xi^{i'_k},
\label{558zaz}
\eea
where
\bea
\hat\varphi_{i'_1\dots i'_k} &=&
\left[\nabla^g_{(i_1}\cdots\nabla^g_{i_k)}
e^{-2\zeta}g^{-1/2}\varphi\right](x')
\\
&=&
\left[
\left(\nabla^g_{(i_1}-2\zeta_{;(i_1}\right)\cdots
\left(\nabla^g_{i_{2k})}-2\zeta_{;i_{2k})}\right)
(g^{-1/2}\varphi)\right](x').
\nonumber
\eea
and compute the Gaussian average over $\xi$
to get the result.  

Notice that the diagonal values of the derivatives of the
function $\zeta$, and, therefore,
the coefficients $\hat\varphi_{i_1\dots i_k}$ and $c_k$,
are polynomial in the derivatives of the curvature of the metric $g$.
By using (\ref{511mmm}) we obtain,
in particular,
\bea
\hat\varphi &=& g^{-1/2}[\varphi],
\label{520via}
\\
\hat\varphi_i &=& [\nabla^g_i(g^{-1/2}\varphi)],
\label{473naa}\\
\hat\varphi_{ij} &=& [\nabla^g_{(i}\nabla^g_{j)} (g^{-1/2}\varphi)]
-2[\zeta_{;ij}g^{-1/2}\varphi],
\label{474naa}
\\
\hat\varphi_{ijk} &=& [\nabla^g_{(i}\nabla^g_{j}\nabla^g_{k)} (g^{-1/2}\varphi)]
-6[\zeta_{(ij}\nabla^g_{k)}(g^{-1/2}\varphi)]
-2[\zeta_{;(ijk)}g^{-1/2}\varphi],
\label{474naab}
\\
\hat\varphi_{ijkl} &=& [\nabla^g_{(i}\nabla^g_{j}\nabla^g_{k}\nabla^g_{l)} (g^{-1/2}\varphi)]
-12[\zeta_{;(ij}\nabla^g_{k}\nabla^g_{l)}(g^{-1/2}\varphi)]
-8[\zeta_{;(ijk}\nabla^g_{l)}(g^{-1/2}\varphi)]
\nonumber\\
&&+\left(12[\zeta_{;(ij}\zeta_{;kl)}]
-2[\zeta_{;(ijkl)}]\right) g^{-1/2}[\varphi].
\label{474via}
\eea
Finally, by using the  diagonal values of the derivatives of the
function $\zeta$, (\ref{511mmm}), and (\ref{512mmm}), 
(\ref{514via}), we obtain the coefficients $c_0$, $c_1$ and $c_2$.
Of course, in the case of the flat metric we recover the
earlier result (\ref{55via}).


\subsection{Morse Lemma}

We say that a smooth real valued symmetric  
function $\Sigma(x,x')$ on $M\times M$ has a 
{\it non-degenerate critical
point on the diagonal} if:
\begin{enumerate}
\item 
the first derivatives vanish on the diagonal, 
$
[\Sigma_{,i}]=0,
$
and
\item
the Hessian  is positive definite on the diagonal,
$G_{ij}=[\Sigma_{,ij}]>0.$

\end{enumerate}

\begin{lemma}
\label{lemma0}
Let $U$ be a sufficiently small open set in a 
manifold 
$M$, $\Sigma: U\times U\to\RR $ be a smooth real valued symmetric non-negative 
function that has a non-degenerate critical point on the diagonal
and vanishes on the diagonal, that is, $[\Sigma]=[\Sigma_{,i}]=0$
and $G_{ij}=[\Sigma_{,ij}]>0$.
Then there exists a local diffeomorphism $\eta^{a}=\eta^{a}(x,x')$
such that the function $\Sigma$ has the form
\be
\Sigma(x,x')=\frac{1}{2}G_{ab}(x')\eta^{a}(x,x')\eta^{b}(x,x').
\label{513qq}
\ee

\end{lemma}

\noindent
{\it Proof.}
We pick some metric $g_{ij}$ and define the corresponding 
Ruse-Synge function $\sigma(x,x')$ and the variables 
$\xi^{i'}=-g^{i'j'}\sigma_{j'}$ introduced in
(\ref{514xxz}).
We expand the function $\Sigma$ in the covariant Taylor series
\be
\Sigma(x,x')=
\sum_{k=2}^\infty\frac{1}{k!}
\Sigma_{i'_1\dots i'_k}
\xi^{i'_1}\cdots
\xi^{i'_k}\,,
\label{525zza}
\ee
where $\Sigma_{i'_1\dots i'_k}=
\left[\Sigma_{;(i_1\dots i_k)}\right](x')$.
This can be written in the form
\be
\Sigma(x,x')=\frac{1}{2}A_{i'j'}(x,x')\xi^{i'}\xi^{j'},
\ee
where
\be
A_{i'j'}(x,x')=
\sum_{k=0}^\infty\frac{2}{(k+2)!}
\Sigma_{i'j'i'_1\dots i'_k}
\xi^{i'_1}\cdots
\xi^{i'_k}\,.
\ee
The matrix $A$ is real and symmetric, so it can be written 
(nonuniquely) in the form
$A=B^THB$;
that is,
\be
A_{i'j'}=G_{ab}B^{a}{}_{i'} B^{b}{}_{j'}.
\ee
The matrix $B$ is defined up to an orthogonal matrix,
that is, up to a transformation
$B\mapsto UB$ 
with the matrix $U$ satisfying $U^TGU=G$.
Then the function $\Sigma$ takes the Morse form (\ref{513qq})
with
$
\eta^{a}=B^{a}{}_{i'}\xi^{i'}.
$
The Morse diffeomorphism is obviously also
defined up to an orthogonal
 transformation $\eta\mapsto U\eta$.

The Morse diffeomorphism $\eta^{a}=\eta^{a}(x,x')$
can be computed explicitly in terms
of the Taylor series
\be
\eta^{a}=
\sum_{k=1}^\infty 
\frac{1}{k!}
\eta^a{}_{i'_1\dots i'_k}\xi^{i'_1}\cdots \xi^{i'_k};
\ee
the coefficients 
$\eta^a{}_{i'_1\dots i'_k}
=[\eta^a{}_{;(i_1\dots i_k)}](x')$
here will be 
expressed in terms of the derivatives 
$\Sigma_{i'_1\dots i'_k}$
of the function $\Sigma$
on the diagonal at the point $x'$.
They can be obtained by substituting this Taylor series in (\ref{513qq})
and comparing it with (\ref{525zza}).
The solution is {\it not unique}.
The first coefficient can be chosen to be a frame of 
vectors $\eta^a{}_{i'}$ 
at the point $x'$ determined by
\be
G_{ab}\eta^a{}_{i'}\eta^b{}_{j'}=\Sigma_{i'j'}.
\ee
Of course, the matrix $\eta^a{}_{i'}$ is defined up to an orthogonal
transformation.


\subsection{Asymptotics of Laplace Type Integrals}

We will need the following lemma.  We fix some metric 
$g_{ij}$; all covariant derivatives and the curvature are defined with respect 
to this metric.

\begin{lemma}
\label{lemma2}
Let $x'$ be a point in a manifold $M$ and $U$ be a sufficiently small 
neighborhood of this point. Let $\Sigma: U\times U\to\RR $ be a smooth real 
valued symmetric non-negative function that has a non-degenerate critical 
point on the diagonal and vanishes on the diagonal. 
Let $\varphi(x,x')$ be a smooth scalar density of weight $1$
and 
\be
F(\varepsilon)=(4\pi\varepsilon)^{-n/2}
\int\limits_{U}dx\;\exp\left(-\frac{1}{2\varepsilon}\Sigma(x,x')\right)
\varphi(x,x').
\label{541zza}
\ee
Then there is the asymptotic expansion
as $\varepsilon\to 0^+$
\be
F(\varepsilon)
\sim 
\sum_{k=0}^\infty \varepsilon^{k} F_k.
\label{531via}
\ee
The coefficients $F_k$ do not depend on the domain $U$; they depend only on the 
derivatives of the functions $\varphi$ and $\Sigma$ at the point $x'$.

Let $\Sigma_{i'_1\dots i'_k} 
=[\nabla^g_{(i_1}\cdots\nabla^g_{i_k)}\Sigma](x')$
be the symmetrized covariant derivatives of the function 
$\Sigma$ on the diagonal at the point $x'$, in particular,
let $G_{ij}=[\Sigma_{,ij}](x')$ be the Hessian on the diagonal, 
$G^{ij}$ be the inverse 
of this matrix and $G=\det G_{ij}$ be its determinant.
Let 
$\varphi_{i'_1\dots i'_k}
=[\nabla^g_{(i_1}\cdots\nabla^g_{i_k)}g^{-1/2}\varphi](x')$ be the 
symmetrized covariant derivatives of
the function $\varphi$ on the diagonal at the point $x'$.
Then:
\begin{enumerate}
\item
the coefficients $F_k$ have the form
$F_k=G^{-1/2}\tilde F_k$, where
\item
$\tilde F_k$ are linear in the derivatives, $\varphi_{i'_1\dots i'_k}$, of the 
function $\varphi$ at the point $x'$ and
\item
polynomial in the inverse Hessian, $G^{ij}$, and 
the derivatives, $\Sigma_{i'_1\dots i'_k}$, $k>2$, of the function 
$\Sigma$ on the diagonal at $x'$ of order higher than $2$.

\item
The first two coefficients are
\bea
F_0 &=&G^{-1/2}[\varphi],
\label{532via}
\\
F_1 &=& G^{-1/2}g^{1/2}\Biggl\{G^{ij}[\nabla^g_{i}\nabla^g_{j}(g^{-1/2}\varphi)]
-G^{ij}G^{pq}\Sigma_{ipq}[\nabla^g_j(g^{-1/2}\varphi)]
\\
&&
+\Biggl[
-\frac{1}{3}G^{ij}R^g_{ij}
+\frac{1}{12}\Bigl(2G^{il}G^{jm}
+3G^{ij}G^{lm}
\Bigr)G^{kn}\Sigma_{ijk}\Sigma_{lmn}
\nonumber\\
&&-\frac{1}{4}G^{ij}G^{kl}\Sigma_{ijkl}
\Biggr]
g^{-1/2}[\varphi]\Biggr\}.
\label{533via}
\eea
\item
In the case when the function $\varphi$ and its first derivative vanish
on the diagonal, $[\varphi]=[\nabla^g_i\varphi]=0$, the third coefficient is
\bea
F_2 &=& g^{1/2}G^{-1/2}\Biggl\{
\frac{1}{2}G^{ij}G^{kl}[\nabla^g_{(i}\nabla^g_j\nabla^g_k\nabla^g_{l)}g^{-1/2}\varphi]
\label{533viax}\\
&&
-\frac{1}{3}\left(
2G^{ij}G^{qk}
+3G^{iq}G^{jk}
\right)G^{pl}\Sigma_{ipq}
[\nabla^g_{(j}\nabla^g_k\nabla^g_{l)}g^{-1/2}\varphi]
\nonumber
\\
&&
+\Biggl[
-\frac{1}{3}\left(G^{ij}G^{kl}+2G^{ik}G^{jl}\right)R^g_{ij}
-\frac{1}{4}\left(G^{pq}G^{kl}
+4G^{kp}G^{lq}\right)G^{pq}\Sigma_{ijpq}
\nonumber\\
&&
+\frac{1}{72}\Biggl(2G^{ij}G^{pr}G^{qs}G^{kl}
+3G^{ij}G^{pq}G^{rs}G^{kl}
+6G^{ik}G^{jl}G^{pq}G^{rs}
\nonumber\\
&&
+12G^{ij}G^{pq}G^{kr}G^{ls}
+12G^{ij}G^{pr}G^{kq}G^{sl}
\Biggr)\Sigma_{ipq}\Sigma_{jrs}
\Biggr][\nabla^g_{(k}\nabla^g_{l)}g^{-1/2}\varphi]
\Biggr\}.
\nonumber
\eea
\end{enumerate}

\end{lemma}

\noindent
{\it Remark.} Notice that the metric $g$ is arbitrary, in particular,
it could be taken to be equal to the Hessian $g_{ij}=G_{ij}$. 

\noindent
{\it Proof.}
First, it is easy to show that the asymptotic expansion does not depend on the 
size of the domain $U$; so, it can be assumed to be sufficiently small. Then 
for 
a sufficiently small $U$ there is a Morse diffeomorphism $\eta^a=\eta^a(x,x')$ 
so that $\Sigma=\frac{1}{2}G_{ab}\eta^a\eta^b$. Then the integral takes the form
\be
F(\varepsilon)=(4\pi\varepsilon)^{-n/2}
\int\limits_{\tilde U}d\eta\;
\exp\left(-\frac{1}{4\varepsilon}\left<\eta,G\eta\right>\right)
f(\eta),
\ee
where 
$\tilde U$ is the corresponding domain for the variables $\eta$
and
\be
f(\eta)=
\left(\det\left(\frac{\partial \eta^a}{\partial x^i}\right)\right)^{-1}
\varphi(x(\eta)).
\ee
Now, by applying Lemma \ref{lemma1} we get the asymptotic expansion
\be
F(\varepsilon)\sim
\sum_{k=0}^\infty \varepsilon^{k}F_k,
\ee
where $F_k=\frac{1}{k!}G^{-1/2}(\Delta^\eta_G f)(0)$.

The coefficients $F_k$ can be computed 
explicitly now by using the Taylor series.
We decompose the function $\Sigma$ via
\be
\Sigma=\frac{1}{2}G_{i'j'}\xi^{i'}\xi^{j'}+\hat \Sigma,
\ee
where $\xi^{i'}$ are the variables introduced in
(\ref{514xxz}) and
\be
\hat \Sigma(\xi)=\sum_{k=3}^\infty \frac{1}{k!}
\Sigma_{i'_1\dots i'_k}\xi^{i'_1}\cdots \xi^{i'_k}.
\label{573zaz}
\ee
where $\Sigma_{i'_1\dots i'_k}=[\Sigma_{;(i_1\dots i_k)}](x')$.
Then by changing the variables $x\mapsto \xi$
and using (\ref{59qqq})
the integral takes the form
\be
F(\varepsilon)=(4\pi\varepsilon)^{-n/2}
\int\limits_{\hat U}d\xi\; g^{1/2}(x')
\exp\left(-\frac{1}{4\varepsilon}\left<\xi,G\xi\right>\right)
\psi(\xi,\varepsilon),
\label{541zzb}
\ee
where $\hat U$ is the modified domain and 
\be
\psi(\xi,\varepsilon)=
\exp\left(-\frac{1}{2\varepsilon}\hat\Sigma(\xi)\right)
\hat\varphi(\xi),
\ee
with $\hat\varphi(\xi)$
defined by (\ref{557ccx}).

Next, by rescaling the variables $\xi^{i'}\mapsto 
\sqrt{\varepsilon}\,\xi^{i'}$, 
we extend the integration to the whole space $\RR^n$ so that
the asymptotics of the integral is given by the Gaussian average
\be
F(\varepsilon)\sim g^{1/2}G^{-1/2}
\left<\psi(\sqrt{\varepsilon}\xi,\varepsilon)\right>_{G}.
\ee
Now, we expand the function $\psi(\sqrt{\varepsilon}\xi,\varepsilon)$ 
in powers of $\varepsilon$
\be
\psi(\sqrt{\varepsilon}\xi,\varepsilon)=
\sum_{k=0}^\infty \varepsilon^{k/2}
\psi_{k/2}(\xi),
\ee 
where $\psi_{k/2}(\xi)$ are polynomials in $\xi$.
It is easy to see that the half-integer order 
coefficients $\psi_{k+1/2}(\xi)$ are odd polynomials
and the integer order coefficients $\psi_{k}(\xi)$ 
are even polynomials. Therefore, the 
Gaussian average of the half-integer order
coefficients vanish, $\left<\psi_{k+1/2}(\xi)\right>_G=0$. 
Thus, finally we obtain the asymptotic expansion (\ref{531via}) with
only integer powers of $\varepsilon$ with
\be
F_k=g^{1/2}G^{-1/2}
\left<\psi_{k}(\xi)\right>_G.
\ee
Then by computing the Gaussian average (\ref{53via})
we get the explicit form of the coefficients of the 
asymptotic expansion.

By 
using the covariant Taylor expansions of the function $\hat\varphi$, 
(\ref{558zaz}), and of the function $\hat\Sigma$, (\ref{573zaz}), we obtain 
\bea
\psi_0(\xi) &=& \hat\varphi,
\\
\psi_1(\xi) &=& \frac{1}{2}\hat\varphi_{i'j'}\xi^{i'}\xi^{j'}
-\frac{1}{48}\left(
4\Sigma_{i'j'k'}\hat\varphi_{l'}
+\Sigma_{i'j'k'l'}\hat\varphi
\right)
\xi^{i'}\xi^{j'}\xi^{k'}\xi^{l'}
\nonumber\\
&&
+\frac{1}{288}\Sigma_{i'j'k'}\Sigma_{l'm'n'}\xi^{i'}\xi^{j'}\xi^{k'}\xi^{l'}
\xi^{m'}\xi^{n'}\hat\varphi,
\eea
where $\hat\varphi$, $\hat\varphi_i$ and $\hat\varphi_{ij}$ are
given by (\ref{520via})-(\ref{474naa}). 

By computing the Gaussian average this gives
\bea
F_0 &=& g^{1/2}G^{-1/2}\hat\varphi,
\\
F_1 &=& g^{1/2}G^{-1/2}\Biggl\{
G^{ij}\hat\varphi_{ij}
-G^{ij}G^{kl}\Sigma_{(ijk}\hat\varphi_{l)}
-\frac{1}{4}G^{ij}G^{kl}\Sigma_{ijkl}\hat\varphi
\nonumber\\
&&
+\frac{5}{12}G^{ij}G^{kl}G^{mn}\Sigma_{(ijk}\Sigma_{lmn)}\hat\varphi\Biggr\}.
\eea
By using Lemma 2.1 of \cite{novoseltsev05}
we get
\be
G^{ij}G^{kl}G^{mn}\Sigma_{(ijk}\Sigma_{lmn)}
=\frac{1}{5}G^{ij}G^{kl}G^{mn}\left(
2\Sigma_{ikm}\Sigma_{jln}
+3\Sigma_{ijm}\Sigma_{kln}
\right).
\label{559via}
\ee
Therefore,
\bea
F_1 &=& g^{1/2}G^{-1/2}\Biggl\{
G^{ij}\hat\varphi_{ij}
-G^{ij}G^{kl}\Sigma_{ijk}\hat\varphi_{l}
-\frac{1}{4}G^{ij}G^{kl}\Sigma_{ijkl}\hat\varphi
\nonumber\\
&&
+\frac{1}{12}
\left(2G^{il}G^{jm}
+3G^{ij}G^{lm}\right)G^{kn}\Sigma_{ijk}\Sigma_{lmn}
\hat\varphi\Biggr\}.
\eea
Finally, by using eqs. (\ref{520via})-(\ref{474naa})
we get the result (\ref{533via}).

In the case when the function $\varphi$ and its first derivative vanish
on the diagonal we also get
\bea
\psi_2(\xi) &=& \frac{1}{24}
\hat\varphi_{i'j'k'l'}\xi^{i'}\xi^{j'}\xi^{k'}\xi^{l'}
-\frac{1}{72}\Sigma_{i'j'k'}\hat\varphi_{l'm'n'}\xi^{i'}\xi^{j'}\xi^{k'}\xi^{l'}\xi^{m'}\xi^{n'}
\\
&&
-\frac{1}{96}\Sigma_{i'j'k'l'}\hat\varphi_{m'n'}\xi^{i'}\xi^{j'}\xi^{k'}\xi^{l'}\xi^{m'}\xi^{n'}
+\frac{1}{576}\Sigma_{i'j'k'}\Sigma_{l'm'n'}\hat\varphi_{p'q'}
\xi^{i'}\xi^{j'}\xi^{k'}\xi^{l'}\xi^{m'}\xi^{n'}
\xi^{p'}\xi^{q'}.
\nonumber
\eea
By computing the Gaussian average over $\xi$ we get
\bea
F_2 &=& g^{1/2}G^{-1/2}\Biggl\{
\frac{1}{2}G^{ij}G^{kl}\hat\varphi_{ijkl}
-\frac{5}{3}G^{ij}G^{kl}G^{mn}\Sigma_{(ijk}\hat\varphi_{lmn)}
\\
&&
-\frac{5}{4}G^{ij}G^{kl}G^{mn}\Sigma_{(ijkl}\hat\varphi_{mn)}
+\frac{35}{72}G^{ij}G^{kl}G^{mn}G^{pq}\Sigma_{(ijk}\Sigma_{lmn}\hat\varphi_{pq)}.
\Biggr\}
\eea
We use Lemma 2.1 of \cite{novoseltsev05}
to get
\bea
G^{ij}G^{kl}G^{mn}\Sigma_{(ijkl}\hat\varphi_{mn)}
&=&\frac{1}{5}G^{ij}G^{kl}G^{mn}\left(\Sigma_{ijkl}\hat\varphi_{mn}
+4\Sigma_{ikmn}\hat\varphi_{jl}\right)
\\
G^{ij}G^{kl}G^{mn}G^{pq}\Sigma_{(ijk}\Sigma_{lmn}\hat\varphi_{pq)}
&=&
\frac{1}{35}
G^{ij}G^{kl}G^{mn}G^{pq}
\Biggl(
2\Sigma_{ikm}\Sigma_{jln}\hat\varphi_{pq}
+3\Sigma_{ijk}\Sigma_{lmn}\hat\varphi_{pq}
\nonumber\\
&&
+6\Sigma_{ijk}\Sigma_{mnp}\hat\varphi_{lq}
+12\Sigma_{ijk}\Sigma_{lmp}\hat\varphi_{nq}
+12\Sigma_{ikm}\Sigma_{jlp}\hat\varphi_{nq}
\Biggr).
\nonumber\\
\eea
Now, by using these equations together with (\ref{559via}) we obtain
\bea
F_2 &=& g^{1/2}G^{-1/2}\Biggl\{
\frac{1}{2}G^{ij}G^{kl}\hat\varphi_{ijkl}
-\frac{1}{3}\left(2
G^{ij}G^{qk}
+3G^{iq}G^{jk}
\right)G^{pl}\Sigma_{ipq}\hat\varphi_{jkl}
\\
&&
+\Biggl[
-\frac{1}{4}\left(G^{pq}G^{kl}
+4G^{kp}G^{lq}\right)G^{pq}\Sigma_{ijpq}
\nonumber\\
&&
+\frac{1}{72}\Biggl(2G^{ij}G^{pr}G^{qs}G^{kl}
+3G^{ij}G^{pq}G^{rs}G^{kl}
+6G^{ik}G^{jl}G^{pq}G^{rs}
\nonumber\\
&&
+12G^{ij}G^{pq}G^{kr}G^{ls}
+12G^{ij}G^{pr}G^{kq}G^{sl}
\Biggr)\Sigma_{ipq}\Sigma_{jrs}
\Biggr]\hat\varphi_{kl}
\Biggr\}.
\nonumber
\eea
Finally, by using eqs. (\ref{520via})-(\ref{474via}) and
(\ref{511mmm})-(\ref{514via})
we obtain
(\ref{533viax}).

Of course, in the particular case when $\Sigma$ is the Ruse-Synge 
function, $\Sigma=\sigma^g$ of the metric $g$, 
all its symmetrized covariant
derivatives of order higher than two 
vanish on the diagonal and the second derivative is equal to the metric, 
which gives 
the earlier result (\ref{555zaz})-(\ref{522viax}).



\section{Asymptotics of Heat Traces}
\setcounter{equation}0

\subsection{Classical Heat Trace}
 
First of all, it is easy to see that 
the asymptotics of the classical heat trace as $t\to \infty$ 
are determined by the bottom eigenvalue
\be
\Theta_{\pm}(t) \sim d^\pm_1\exp\left(-t\lambda_1^\pm\right),
\ee
where $d^\pm_1=\tr P^\pm_1$ is the multiplicity of the first eigenvalue
and $P^\pm_1$ is the projection to the first eigenspace.
We will be primarily interested in the asymptotics as $t\to 0$.

For Laplace type operators $L_\pm$ 
there is an asymptotic expansion of the heat kernel
$U_\pm(t;x,x')$ 
in the neighborhood of the diagonal 
as $t\to 0$
(see e.g. \cite{avramidi91,avramidi00,avramidi10,avramidi15})
\bea
U_\pm(t;x,x')\sim 
(4\pi)^{-n/2}\exp\left(-\frac{\sigma_\pm}{2t}\right)
\sum_{k=0}^\infty  t^{k-n/2} \tilde a^\pm_k(x,x'),
\label{513xxcd}
\eea
where 
\bea
\tilde a^\pm_k(x,x')
&=& \frac{(-1)^k}{k!}M_\pm^{1/2}(x,x')\cP_\pm(x,x')
a^\pm_k(x,x')
\nonumber\\
&=&
\frac{(-1)^k}{k!}g_\pm^{1/4}(x)g_\pm^{1/4}(x')
e^{\zeta_\pm(x,x')}\cP_\pm(x,x')
a^\pm_k(x,x'),
\label{513xxcq}
\eea
$\sigma_\pm=\sigma_\pm(x,x')$ is the   Ruse-Synge function of the metric 
$g_\pm$, $M_\pm=M_\pm(x,x')$ is the Van Vleck-Morette determinant, 
$\cP_\pm=\cP_\pm(x,x')$ is the operator of parallel transport of sections along 
the geodesic in the connection $\nabla^\pm$ and the metric $g_\pm$ from the 
point $x'$ to the point $x$ and $a^\pm_k=a^\pm_k(x,x')$ are the usual heat 
kernel coefficients. In particular,
\cite{avramidi00}
\be
a^\pm_0 = I,
\label{64via}
\ee
and
\be
[a^\pm_1] = Q_\pm-\frac{1}{6}R_\pm I,
\label{65via}
\ee
where $R_\pm$ is the scalar curvature of the metric $g_\pm$
and for the Dirac operator $Q_\pm$ is given by
(\ref{qsxxa}).

Therefore, there is the asymptotic expansion (\ref{513xxc})
of the classical heat trace (\ref{532xxc}).
This is the classical heat 
trace asymptotics of Laplace type operators.
By using the off-diagonal expansion of the heat kernel (\ref{513xxcd}) for the 
Laplace type operator (and using the diagonal values of the derivatives of the 
functions $\sigma_\pm, M_\pm, \cP_\pm$) one can also obtain the asymptotic 
expansion of the classical spectral 
invariant $H(t)$, (\ref{12via}), for the Dirac type operator,
\be
H_\pm(t)\sim (4\pi)^{-n/2}\sum_{k=0}^\infty t^{k-n/2} H_k^\pm,
\ee
where
\be
H_k^\pm =\frac{(-1)^k}{k!}\int\limits_M dx\; g_\pm^{1/2}\tr\; [D_\pm a^\pm_k].
\ee
Here we used, in particular, a useful relation
\be
[D_\pm \tilde a_k]=\frac{(-1)^k}{k!}g_\pm^{1/2}[D_\pm a_k]
\label{68via}
\ee
Further, by using 
\be
[D_\pm \cP_\pm]= S_\pm
\label{68viab}
\ee
and the results of \cite{avramidi91,avramidi00,avramidi15} 
we get 
\bea
H_0^\pm &=& \int\limits_M dx\; g_\pm^{1/2}\tr\; S_\pm,
\\
H_1^\pm &=& \int\limits_M dx\; g_\pm^{1/2}
\tr\left\{i\gamma^j_\pm\left(-\frac{1}{2}\nabla^\pm_jQ_\pm
-\frac{1}{6}\nabla^\pm_k\cR_\pm^k{}_j\right)
+S_\pm\left(\frac{1}{6}R_\pm-Q_\pm\right)
\right\}.
\label{69via}
\eea
For the manifolds without boundary we can safely neglect the
total derivative terms in (\ref{69via}).


\subsection{Combined Heat Trace for Laplace Type Operators}

To compute the asymptotics of the relative spectral invariants 
$\Psi(t,s)$ and $\Phi(t,s)$
we 
rescale the variables $t \mapsto \varepsilon t$ and $s\mapsto \varepsilon s$ 
and study the asymptotics as $\varepsilon\to 0$
or $\varepsilon\to\infty$. 
It is easy to see that 
the asymptotics of the combined heat traces $X(t,s)$, (\ref{238ccx}),
and $Y(t,s)$, (\ref{534xxc}),
as $\varepsilon\to \infty$ are 
determined by the bottom eigenvalues $\lambda^\pm_1$
and $\mu^\pm_1$,
\bea
X\left(\varepsilon{}t,\varepsilon{}s\right)
&\sim & \exp\left[-\varepsilon{}\left(t\lambda_1^++s\lambda_1^-\right)\right]
\Tr P_1^+P_1^-,
\\
Y\left(\varepsilon{}t,\varepsilon{}s\right)
&\sim & \exp\left[-\varepsilon{}
\left(t(\mu_1^+)^2+s(\mu_1^-)^2\right)\right]\;\mu_1^+\mu_1^-\;
\Tr P_1^+ P_1^-.
\eea
We will be interested mainly in the asymptotics as $\varepsilon\to 0$.
In this subsection we prove the Theorem \ref{theorem1}
for the combined heat trace $X(t,s)$.

{\bf Proof of Theorem \ref{theorem1}, Part I.}
The combined trace $X(t,s)$ is
given by the integral (\ref{533xxca}) 
over $M\times M$
of the form
\bea
X(t,s) &=& \int\limits_{M\times M} dx\; dx' f_1(t,s;x,x'),
\eea
where
\be
f_1(t,s;x,x')=\frac{1}{2}\tr\Bigl\{U_+(t,x,x')U_-(s,x',x)
+U_-(s,x,x')U_+(t,x',x)\Bigr\}.
\ee
Notice that we made it manifestly symmetric under the exachange
$(t,L_+)\leftrightarrow (s,L_-)$ by symmetrizing the integrand
in $x$ and $x'$ (alternatively, we could do the symmetrization
$(t,L_+)\leftrightarrow (s,L_-)$
at the end of the calculations).

Let $r^\pm_{\rm inj}$ be the injectivity radii of the metrics $g_\pm$ and
\be
\rho=\min\{r^+_{\rm inj},r^-_{\rm inj}\}.
\ee
We fix a point $x'$ in the manifold $M$
(of course, we could instead fix the point $x$; we have to do it both ways
to achieve the required symmetry $(t,L_+)\leftrightarrow(s,L_-)$ of the heat trace).

Let $B^\pm_r(x')$ be the geodesic balls in the metric $g^\pm_{ij}$ 
centered at $x'$ of radius 
$r<\rho$ smaller than the injectivity radii $r^\pm_{\rm inj}$.
Let $B_r(x')\subset B^+_r(x')\cap B^-_r(x')$ be an open set
contained in both of these balls.
We decompose the combined traces as follows
\bea
X(t,s) &=& X_{\rm diag}(t,s)+X_{\rm off-diag}(t,s),
\eea
where
\bea
X_{\rm diag}(t,s)&=&
\int\limits_{M}dx'\,\int\limits_{B_r(x')}dx\,f_1(t,s;x,x'),
\\
X_{\rm off-diag}(t,s)&=&\int\limits_M dx' \int\limits_{M-B_r(x')}dx\;f_1(t,s;x,x')\,.
\eea

To estimate these integrals we will need the following lemma.

\begin{lemma}
\label{lemma6}
The off-diagonal part $X_{\rm off-diag}(\varepsilon{}t,\varepsilon{}s)$ of the 
combined heat trace is exponentially small as $\varepsilon\to 0$ and does 
not contribute to its asymptotic expansion, that is, as $\varepsilon\to 0$
\be
X(\varepsilon{}t,\varepsilon{}s)\sim X_{\rm 
diag}(\varepsilon{}t,\varepsilon{}s).
\ee
\end{lemma}

\noindent
{\it Proof.}
This can be proved by using the standard elliptic estimates of the heat kernel.
For any $x\in M-B_r(x')$ and $0<t<1$
there is an estimate \cite{grigoryan09}
\be
\left|U_\pm(t;x,x')\right|\le 
C_1 t^{-n/2}\exp\left(-\frac{r^2}{4t}\right),
\ee
where $C_{1}=C_1(r)$ is some constant. 
Therefore,
\be
|f_1(t,s;x,x')|\le 
C_2\varepsilon^{-n}t^{-n/2}s^{-n/2}\exp\left[-\frac{r^2}{4\varepsilon{}}
\left(\frac{1}{t}+\frac{1}{s}\right)\right]
\ee
with some constant $C_2$.
This means that as $\varepsilon\to 0$
\be
X_{\rm off-diag}(\varepsilon{}t,\varepsilon{}s)\sim 0.
\ee
The statement follows.

Now, by using this Lemma \ref{lemma6} and
the asymptotic expansion of the heat kernel (\ref{513xxcd})
we obtain the asymptotic expansion of the combined heat trace 
(\ref{533xxc}) as $\varepsilon\to 0$,
\be
X\left(\varepsilon{}t,\varepsilon{}s\right) \sim 
(4\pi\varepsilon)^{-n/2}
\sum_{m=0}^\infty \varepsilon^{m}
X_{m}(\varepsilon,t,s),
\label{622via}
\ee
where
\bea
X_{m}(\varepsilon,t,s) 
\label{527xxc}
=\left(4\pi\varepsilon ts\right)^{-n/2}
\int\limits_M dx'
\int\limits_{B_r(x')}dx\;
\exp\left\{-\frac{1}{2\varepsilon ts{}}\Sigma(t,s;x,x')\right\}
\Lambda_{m}(t,s;x,x'),
\nonumber\\
\eea
and
\bea
\Sigma(t,s;x,x') &=& s\sigma_+(x,x')
+t\sigma_-(x,x'),
\label{529zzc}
\\[5pt]
\Lambda_{m}(t,s;x,x')
&=& \sum_{j=0}^m
t^{m-j}s^{j}
\frac{1}{2}\tr\,\left\{ 
\tilde a^+_{m-j}(x,x')\tilde a^-_j(x',x)
+\tilde a^-_{m-j}(x,x')\tilde a^+_j(x',x)\right\}.
\nonumber\\
\label{530zzc}
\eea

Next, we compute the asymptotic expansion of 
the function $X_m(\varepsilon,t,s)$.
\begin{lemma}
\label{lemma7}
There is the asymptotic expansion as $\varepsilon\to 0$
\be
X_{m}\left(\varepsilon,t,s\right) \sim
\sum_{k=0}^\infty \varepsilon^{k}
X_{m,k}(t,s),
\label{626via}
\ee
where
\be
X_{m,k}(t,s)=\int\limits_M dx\; 
g^{1/2}(t,s)b_{m,k}(t,s).
\label{631viab}
\ee
The coefficients $b_{m,k}(t,s)$ are scalar invariants constructed 
polynomially from the diagonal values of the derivatives of the function 
$\Lambda_{m}$ and the derivatives of the function $\Sigma(t,s)$ of order 
higher than $2$ as well as the metrics $g_+^{ij}$, $g_-^{ij}$ and 
$g_{ij}(t,s)$, (\ref{113via}). 
The coefficients $b_{m,k}(t,s)$ are homogeneous functions of $t$ and 
$s$ of degree $(m+k)$ and the coefficients $X_{m,k}(t,s)$ are 
homogeneous functions of $t$ and $s$ of degree $(m+k-n/2)$. In particular,
\bea
b_{m,0}(t,s) &=&\sum_{j=0}^m
\frac{(-1)^{m}}{(m-j)!j!}t^{m-j}s^{j}
\tr \left\{[a_{m-j}^+]\,[a_j^-]\right\}.
\label{628via}
\eea
\end{lemma}

{\it Proof.}
The function $\Sigma(t,s)=\Sigma(t,s;x,x')$, (\ref{529zzc}), is smooth and 
positive
(here and below we omit the space variables $x$ and $x'$).
It has the absolute minimum on the diagonal, at $x=x'$, equal to 
zero, 
\be
[\Sigma(t,s)]=0
\ee
and has a non-degenerate critical point on the diagonal, that is,
the first derivatives vanish on the diagonal
\be
[\Sigma_{,i}(t,s)]=0,
\ee
and the Hessian
$
[\Sigma_{,ij}(t,s)]
=G_{ij}(t,s),
$
which is exactly equal to the matrix $G_{ij}(t,s)$, defined by (\ref{126via}),
is positive on the diagonal.
Also, by using (\ref{128viab}) we can see that 
the determinant of the Hessian has the form
\be
G = \det G_{ij} = \frac{g_+g_-}{g}.
\label{631via}
\ee
where $g=\det g_{ij}$ and $g_\pm=\det g^\pm_{ij}$. 

Now, by using Lemma \ref{lemma2} we compute the 
asymptotic expansion of the integral (\ref{527xxc})
which gives (\ref{626via}) and proves the first part of the
lemma.
The coefficient $b_{m,0}(t,s)$ is given by (\ref{532via}),
so
\be
b_{m,0} = g^{-1/2}G^{-1/2}[\Lambda_m]
= g_+^{-1/2}g_-^{-1/2}[\Lambda_m].
\label{631viac}
\ee
Further, we compute the diagonal value of the functions $\Lambda_{m}(t,s)$, 
(\ref{530zzc}) 
\be
[\Lambda_{m}(t,s)]
= g_+^{1/2}g_-^{1/2}\sum_{j=0}^m\frac{(-1)^{m}}{(m-j)!j!} 
t^{m-j}s^{j}
\tr\{ [a_{m-j}^+]\,[a_j^-]\},
\ee
to get (\ref{628via}). This proves Lemma \ref{lemma7}.

Thus, by using (\ref{622via}) and (\ref{626via})
we obtain the asymptotic expansion (\ref{1zaab})
of the combined heat trace $X(t,s)$ with the coefficients
\be
b_k(t,s)=\sum_{j=0}^k
b_{k-j,j}(t,s).
\label{634via}
\ee
This proves Theorem \ref{theorem1} for the trace $X(t,s)$.
Finally, by using the relation (\ref{16via}) and the asymptotic
expansions (\ref{513xxc}) and (\ref{1zaab}) we obtain 
the asymptotic expansion (\ref{120via})
of the relative spectral invariant $\Psi(t,s)$
with the coefficients (\ref{543xxca}).
This proves Corollary \ref{corollary1} for the function $\Psi(t,s)$.

\subsection{Combined Heat Trace for Dirac Type Operators}
 
The case for the Dirac type operators is similar to the Laplace type operators; 
so we will omit some details. 
In this subsection we prove the Theorem \ref{theorem1} for the combined heat trace
$Y(t,s)$.

{\bf Proof of the Theorem \ref{theorem1}, Part II.}
The combined heat trace $Y(t,s)$ is given by the integral (\ref{534xxc})
over $M\times M$
of the form
\bea
Y(t,s) &=& \int\limits_{M\times M} dx\; dx' f_2(t,s;x,x'),
\eea
where
\be
f_2(t,s;x,x')=\frac{1}{2}\tr\Bigl\{D_+U_+(t,x,x')D_-U_-(s,x',x)
+D_-U_-(s,x,x')D_+U_+(t,x',x)\Bigr\},
\ee
where the operators $D_\pm$ act 
on the first space argument of the heat kernels.
The method for 
computing 
this integral is essentially the same as for the Laplace type operators. In 
this 
case we use the estimate for the derivative of the heat kernel: for any $x\in 
M-B_r(x')$ and $0<t<1$
\be
\left|D_\pm U_\pm(t;x,x')\right|\le C_3 
t^{-1-n/2}\exp\left(-\frac{r^2}{4t}\right),
\ee
where $C_3$ is some constant. 
By using this estimate it is easy to see that
the off-diagonal part of the 
integral
is exponentially small and
does not contribute to the asymptotic expansion of the trace
$Y(\varepsilon{}t,\varepsilon{}s)$ as $\varepsilon\to 0$.

Therefore, we obtain
\be
Y(\varepsilon{}t,\varepsilon{}s) \sim 
(4\pi\varepsilon)^{-n/2}
\sum_{m=0}^\infty \varepsilon^{m}\tilde Y_{m}(\varepsilon,t,s),
\label{636via}
\ee
where
\bea
\tilde Y_{m}(\varepsilon,t,s)
\label{642aabc}
=(4\pi\varepsilon ts)^{-n/2}
\int\limits_M dx'\int\limits_{B_r(x')}dx\;
\exp\left\{-\frac{1}{2\varepsilon ts{}}\Sigma(t,s;x,x')\right\}
\tilde N_{m}(\varepsilon,t,s,x,x'),
\nonumber\\
\eea
with
\bea
&&
\tilde N_{m}(\varepsilon,t,s,x,x') 
\label{6438via}
\\
&&=
\frac{1}{2}\sum_{j=0}^m t^{m-j} s^{j}\tr\Biggl\{
\left( D_+ -\frac{1}{2\varepsilon{}t}\nu_+(x,x')\right)\tilde a^+_{m-j}(x,x')
\left( D_- -\frac{1}{2\varepsilon{}s}\nu_-(x',x)\right)
\tilde a^-_j(x',x)
\Biggr\}
\nonumber\\
&&
+ x\leftrightarrow x',
\nonumber
\eea
and 
\bea
\nu_\pm(x,x') &=& i\gamma^j_\pm(x)\sigma^\pm_{,j}(x,x').
\label{639via}
\eea
To avoid confusion we stress here once again that the operators $D_\pm$
act on the first space argument of the coefficients $\tilde a^\pm_k$.
The functions $\tilde N_{m}$ depend on $\varepsilon$ in the following way
\bea
\tilde N_{m} =
\frac{1}{\varepsilon^2}N^{(0)}_{m}
+\frac{1}{\varepsilon{}}N^{(1)}_{m}
+N^{(2)}_{m},
\eea
where
\bea
N^{(0)}_{m}&=& \frac{1}{8}(ts)^{-1}\sum_{j=0}^mt^{m-j} s^{j}
\tr\left\{\nu_+(x,x')
\tilde a_{m-j}^+(x,x')\nu_-(x',x)\tilde a_j^-(x',x)\right\}
+ x\leftrightarrow x',
\nonumber\\
\label{642via}
\\
N^{(1)}_{m}&=& -\frac{1}{4}(ts)^{-1}\sum_{j=0}^m t^{m-j} s^{j}\tr\Biggl\{
s\nu_+(x,x')\tilde a_{m-j}^+(x,x')
(D_-\tilde a_j^-(x',x))
\nonumber\\
&&
+t(D_+\tilde a_{m-j}^+(x,x'))\nu_-(x',x)\tilde a_j^-(x',x)
\Biggr\}+ x\leftrightarrow x',
\label{645viax}
\\
N^{(2)}_{m}&=& \frac{1}{2}\sum_{j=0}^m t^{m-j} s^{j}
\tr\left\{(D_+\tilde a_{m-j}^+(x,x'))(D_-\tilde a_j^-(x',x))\right\}
+ x\leftrightarrow x'.
\eea
Therefore, 
\be
\tilde Y_{m}
= \frac{1}{\varepsilon^2}  Y^{(0)}_{m}
+\frac{1}{\varepsilon}  Y^{(1)}_{m}
+Y_{m}^{(2)},
\label{642viax}
\ee
with the obvious notation
\bea
Y^{(i)}_{m}(\varepsilon,t,s)
=(4\pi\varepsilon ts)^{-n/2}
\int\limits_M dx'\int\limits_{B_r(x')}dx\;
\exp\left\{-\frac{1}{2\varepsilon ts{}}\Sigma(t,s;x,x')\right\}
N^{(i)}_{m}(t,s,x,x').
\label{642aab}
\nonumber\\
\eea
Therefore, by using (\ref{642viax}) we have
\be
Y(\varepsilon{}t,\varepsilon{}s) 
\sim 
(4\pi\varepsilon)^{-n/2}
\sum_{m=0}^\infty\varepsilon^{m-2} Y_m(\varepsilon,t,s),
\label{649via}
\ee
where
\bea
Y_0 &=& Y_0^{(0)},
\\
Y_1 &=& Y^{(0)}_1+Y^{(1)}_0,
\eea
and for $m\ge 2$
\be
Y_m =
 Y^{(0)}_{m}
+ Y^{(1)}_{m-1} 
+ Y^{(2)}_{m-2}.
\ee

By using Lemma \ref{lemma2}
again to compute the integral (\ref{642aab}) 
we prove the following lemma.
\begin{lemma}
\label{lemma8}
There is the asymptotic expansion as $\varepsilon\to 0$
\bea
Y^{(i)}_{m}(\varepsilon,t,s) &\sim &
\sum_{k=0}^\infty \varepsilon^{k}
Y^{(i)}_{m,k}(t,s),
\label{650via}
\eea
where
\be
Y^{(i)}_{m,k}(t,s)=\int\limits_M dx\; 
g^{1/2}(t,s) c^{(i)}_{m,k}(t,s).
\ee
The coefficients 
 $ c^{(i)}_{m,k}(t,s)$
are scalars  
constructed polynomially from the diagonal values of the
derivatives of the functions
$N^{(i)}_{m}$ 
and the 
derivatives of the function $\Sigma(t,s)$ 
of order higher than $2$ as well as 
the metric $g_{ij}(t,s)$ and $g_\pm^{ij}$. 
The coefficients  $ c^{(i)}_{m,k}(t,s)$ are
homogeneous functions of $t$ and $s$ 
of degree $(m+k+i-2)$ and the coefficients $Y^{(i)}_{m,k}(t,s)$
are homogeneous functions of $t$ and $s$ of degree 
$(m+k+i-2-n/2)$.
The first coefficients are
\be
c^{(0)}_{m,0}=c^{(1)}_{m,0}=0,
\ee
\be
c^{(2)}_{m,0}(t,s) =
\sum_{j=0}^{m}
\frac{(-1)^{m}}{(m-j)!j!}
t^{m-j}s^{j}
\tr\left\{[D_+ a^+_{m-j}]
[D_- a^-_j]\right\}.
\label{652via}
\ee
\end{lemma}


\noindent
{\it Proof.} The proof of this lemma is essentially the same as 
that of the Lemma \ref{lemma7}.
By using Lemma \ref{lemma2} we compute the 
asymptotic expansion of the integral (\ref{642aab})
which gives (\ref{650via}) and proves the first part of the
lemma.

The coefficient $c^{(i)}_{m,0}(t,s)$ is given by (\ref{532via}),
so
\be
c^{(i)}_{m,0}=g^{-1/2}G^{-1/2}[N^{(i)}_m]
=g_+^{-1/2}g_-^{-1/2}[N^{(i)}_m].
\label{657via}
\ee
Thus, we need to compute the diagonal values of the functions
$N^{(i)}_m$. 
First of all, since the diagonal values of the function
$\sigma_\pm$ and its first derivatives vanish
it is easy to see that the diagonal values of the functions
$\nu_\pm$, (\ref{639via}), vanish,
\be
[\nu_\pm]=0,
\ee
and, therefore,
\be
[N_{m}^{(0)}]=[N_{m}^{(1)}]=0.
\ee
This means that
\be
c^{(0)}_{m,0}=c^{(1)}_{m,0}=0.
\ee

The functions $N^{(2)}_m$ are expressed in terms of the 
coefficients $\tilde a^\pm_k$, which are related to the standard
heat kernel coefficients $a^\pm_k$ by (\ref{513xxcq}). 
Therefore, by using the diagonal values of the 
functions $\sigma_\pm, M_\pm, 
\cP_\pm$, and their derivatives
we obtain 
\be
[N^{(2)}_{m}(t,s)] = g_+^{1/2}g_-^{1/2}
\sum_{j=0}^{m} 
\frac{(-1)^{m}}{(m-j)!j!}t^{m-j}s^j
\tr\left\{[D_+ a^+_{m-j}]
[D_- a^-_j]\right\}.
\ee
Now, by using (\ref{631via}) and (\ref{657via})
we get (\ref{652via}). This proves Lemma \ref{lemma8}.


By using this lemma, we obtain the asymptotic expansion
\bea
Y_{m}(\varepsilon,t,s) &\sim &
\sum_{k=0}^\infty \varepsilon^{k}
Y_{m,k}(t,s),
\label{650viab}
\eea
where
\be
Y_{m,k}(t,s)=\int\limits_M dx\;g^{1/2}(t,s)c_{m,k}(t,s)
\ee
with
\bea
c_{0,k} &=& c_{0,k}^{(0)},
\\
c_{1,k} &=& c^{(0)}_{1,k}+c^{(1)}_{0,k},
\eea
and for $m\ge 2$
\be
c_{m,k} = c^{(0)}_{m,k}
+ c^{(1)}_{m-1,k} 
+ c^{(2)}_{m-2,k}.
\ee
Now, by using (\ref{649via}) and the equations above we obtain 
\be
Y(\varepsilon{}t,\varepsilon{}s) \sim
(4\pi\varepsilon)^{-n/2}
\sum_{k=-1}^\infty \varepsilon^{k-1} 
C_k(t,s),
\label{662via}
\ee
with the coefficients
\be
C_k=\sum_{j=0}^{k+1}
Y_{j, k+1-j}.
\label{662viab}
\ee
Finally, we notice that the first coefficient $C_{-1}$
vanishes since
\be
C_{-1} = Y_{0,0} = Y^{(0)}_{0,0} = 0.
\ee
Thus, we obtain the 
asymptotic expansion (\ref{15zaac}) 
of the combined heat trace $Y(t,s)$ with the coefficients
\be
c_k(t,s)=\sum_{j=0}^{k+1}
c_{j, k+1-j}(t,s).
\label{662vian}
\ee
This proves Theorem \ref{theorem1} for the trace $Y(t,s)$.
Finally, by using the relation (\ref{17via}) and the asymptotic
expansions (\ref{513xxc}) and (\ref{15zaac}) we obtain 
the asymptotic expansion (\ref{121via})
of the relative spectral invariant $\Phi(t,s)$
with the coefficients (\ref{542saax}).
This proves Corollary \ref{corollary1} for the function $\Phi(t,s)$.


\subsection{Specific Cases}

First of all, we notice that since for equal operators $L_-=L_+$ the 
combined trace $X(t,s)$
can be expressed in terms of the classical heat trace
\be
X(t,s) =\Theta(t+s),
\ee
then, by comparing (\ref{1zaab}) and 
(\ref{513xxc}) we see that in this case
\be
B_k(t,s)=(t+s)^{k-n/2} A_k.
\label{674viax}
\ee
Similarly, since for equal operators $D_-=D_+$ the 
combined trace $Y(t,s)$
can be expressed
in terms of the classical heat trace
\be
Y(t,s) = - \partial_t\Theta(t+s),
\ee
then, by comparing (\ref{15zaac}) and 
(\ref{513xxc}) we see that in this case
\be
C_k(t,s)=-\left(k-\frac{n}{2}\right)(t+s)^{k-1-n/2} A_k.
\label{676viax}
\ee
This gives non-trivial relations between the heat kernel coefficients
and their derivatives and provides a useful check of the results.
It is easy to see then that 
for equal operators $L_-=L_+$ and $D_-=D_+$ 
the relative spectral invariants $\Psi(t,s)$
and $\Phi(t,s)$ vanish.

If the Laplace type operators
differ by just a constant,
\be
L_+=L_-+M^2,
\ee
then the metrics and the connections are the same and
\bea
\Theta_+(t) &=& e^{-tM^2}\Theta_-(t),
\\
X(t,s) &=& e^{-tM^2}\Theta_-(t+s),
\eea
and, therefore,
\be
\Psi(t,s)=\left(e^{-tM^2}-1\right)\left(e^{-sM^2}-1\right)\Theta_-(t+s).
\ee
In this case
\bea
B_0(t,s) &=& (t+s)^{-n/2}A_0^-,
\\
B_1(t,s) &=& (t+s)^{1-n/2}A_1^--t(t+s)^{-n/2}M^2A_0^-.
\eea

For the Dirac case suppose that there is an endomorphism $M$
such that it anticommutes with the operator
$D_-$,
\be
D_-M=-MD_-,
\ee
and $M^2$ is a scalar.
Then it is easy to see that
\be
\Tr MD_-\exp(-sD^2_-)=0.
\ee
Now, suppose that
\be
D_+=D_-+M,
\ee
so that (recall that $L_+=D_+^2$)
\be
L_+=L_-+M^2;
\ee
Then it is easy to show that 
\bea
H_+(t) &=& H_-(t)+\Tr M\exp(-tD_-^2),
\\
Y(t,s) &=& -e^{-tM^2}\partial_t\Theta_-(t+s),
\eea
and, hence,
\be
\Phi(t,s)=-\left(e^{-tM^2}-1\right)\left(e^{-sM^2}-1\right)
\partial_t\Theta_-(t+s)
+M^2e^{-(t+s)M^2}\Theta_-(t+s).
\ee
Therefore,
\bea
C_0(t,s) &=& \frac{n}{2}(t+s)^{-1-n/2}A_0^-,
\\
C_1(t,s) &=& \left(\frac{n}{2}-1\right)(t+s)^{-n/2}A_1^-
-\frac{n}{2}t(t+s)^{-1-n/2}M^2A_0^-.
\eea

A more general case is the case of {\it commuting}
operators; then the combined heat traces still simplify significantly,
they can be expressed in terms of the classical one
\bea
X(t,s) &=& \Tr\exp(-tL_+-s L_-),
\label{238ccxa}
\\
Y(t,s) &=& \Tr D_-D_+\exp(-tD^2_+ -sD^2_-).
\eea
Therefore, the asymptotics of the combined traces can be obtained from the 
classical ones.
Notice that the leading symbol of the operators $L=tL_++sL_-$ 
and $L=tD_+^2+sD_-^2$ is determined exactly by
the metric $g^{ij}(t,s)$. Therefore,
in this case the combined traces are given by the
classical trace for the operator $L=tL_++sL_-$.

\section{Explicit Results}
\setcounter{equation}0

\subsection{Laplace Type Operators (Proof of Theorem \ref{theorem2})}

Coming back to the general case, it is easy to see that the first coefficients
are the same as in the commuting case.
The coefficient $B_0$ is obtained by using
(\ref{634via}), (\ref{628via}) and (\ref{64via})
\be
b_0=b_{0,0}=\tr I.
\ee
This proves eq. (\ref{132via}).

The coefficient $b_1$ has the form (by using (\ref{634via}))
\be
b_1=b_{1,0}+b_{0,1}.
\ee
Here the first coefficient is  easy to compute.
By using the well known results (\ref{64via}), (\ref{65via}),
for the coefficients $[a^\pm_k]$
we obtain from (\ref{628via})
\bea
b_{1,0} &=& 
\tr\left\{
-t\,[a_1^+a_0^-]
-s\,[a_0^+a_1^-]
\right\}
\nonumber\\
&=&
\tr\left\{
t\,\left(\frac{1}{6}R_+I-Q_+\right)
+
s\,\left(\frac{1}{6}R_-I-Q_-\right)
\right\}.
\label{72viax}
\eea

The coefficient $b_{0,1}$ 
is determined by the second term of the asymptotics 
(\ref{626via})
of the 
quantity
$X_{0}(\varepsilon t,\varepsilon s)$, (\ref{527xxc}).
By using (\ref{530zzc}), (\ref{513xxcq}), (\ref{631via}) and (\ref{410qqq})
we have
\bea
\Lambda_0 &=& g^{1/2}(x)G^{1/2}(x')
e^{\omega(x,x')}\varphi_1(x,x'),
\eea
where
\be
\omega(x,x') =
\frac{1}{4}\log\left(\frac{g_+(x)}{g(x)}
\frac{g_-(x)}{g(x)}\frac{g(x')}{g_+(x')}\frac{g(x')}{g_-(x')}\right)
+\zeta_+(x,x')+\zeta_-(x,x')
\label{75via}
\ee
and
\be
\varphi_1(x,x')=
\frac{1}{2}\tr\left\{\cP_+(x,x')\cP_-(x',x)
+\cP_-(x,x')\cP_+(x',x)\right\}. 
\ee
By using the fact that $\cP(x',x)=\cP^{-1}(x,x')$ we 
find it convenient to rewrite the function $\varphi_1$
in the form
\be
\varphi_1=
\frac{1}{2}\tr\left(\Pi+\Pi^{-1}\right), 
\ee
where
\be
\Pi(x,x')=\cP_-(x',x)\cP_+(x,x')=\cP_-^{-1}\cP_+.
\ee

Here the function $\zeta_\pm(x,x')$ is defined by (\ref{410qqq}).
Now, by using Lemma \ref{lemma2} 
and eq. (\ref{533via})
we obtain
\bea
b_{0,1}
 &=& ts G^{ij}[\nabla^g_{i}\nabla^g_{j}(e^{\omega}\varphi_1)]
- tsG^{ij}G^{kl}\Sigma_{ijk}[\nabla^g_l(e^{\omega}\varphi_1)]
\nonumber\\
&&
+ts\Biggl(-\frac{1}{3}G^{ij}R^g_{ij}
-\frac{1}{4}G^{ij}G^{kl}\Sigma_{ijkl}
+\frac{1}{6}G^{il}G^{jm}G^{kn}\Sigma_{ijk}\Sigma_{lmn}
\nonumber\\
&&
+\frac{1}{4}G^{ij}G^{lm}G^{kn}\Sigma_{ijk}\Sigma_{lmn}
\Biggr)[\varphi_1],
\eea
where  
$\Sigma_{i_1\dots i_k}=[\nabla^g_{(i_1}\cdots\nabla^g_{i_k)}\Sigma]$
are the coincidence limits of symmetrized  covariant derivatives of $\Sigma$
determined by the metric $g_{ij}$ 
(\ref{113via}).

First of all, we notice that $[\omega]=0$.
We will denote the diagonal values of the derivatives of the function $\omega$
by just adding indices, that is, $\omega_i=[\nabla^g_i\omega]$
and $\omega_{ij}=[\nabla^g_i\nabla^g_j\omega]$. 
By using (\ref{511mmm}) and (\ref{512mmm}) and
the fact that $[\zeta^\pm_{,i}]=0$
we compute
the diagonal values of the first two derivatives
\bea
\omega_{i}&=&W_i,
\\{}
\omega_{ij} &=& \frac{1}{6}R^+_{ij}+\frac{1}{6}R^-_{ij}+W_{ij},
\label{711viaxz}
\eea
where $W_i$ and $W_{ij}$ are defined by (\ref{137zax})
and (\ref{130viaxz}).


Next, it is easy to see that
\be
[\varphi_1]=\tr I, 
\ee
Next, 
since $[\nabla^\pm_i\cP_\pm]=0$ and $[\cP_\pm]=I$ we have
\be
[\nabla^g_i\cP_\pm{}(x,x')] =-[\nabla^g_{i'}\cP_\pm{}(x,x')]
=-\cA^\pm_i,
\label{716viax}
\ee
and
\be
[\nabla^g_i\Pi]=-[\nabla^g_i \Pi^{-1}]=\cC^-_i-\cC^+_i;
\ee
therefore,
\be
[\nabla^g_i\varphi_{1}]=0,
\ee
and 
\be
[\nabla^g_i(e^{\omega}\varphi_1)]
=\omega_{i}\tr I.
\ee
Further, we compute
\bea
[\nabla^g_{i}\nabla^g_{j} (e^{\omega}\varphi_1)]
&=&[\nabla^g_{i}\nabla^g_{j}\varphi_{1}]
+\left(\omega_{ij}+\omega_{i}\omega_{j} \right)\tr I.
\label{713gia}
\eea
By using (\ref{372zaa}) and (\ref{373zaa}) 
we compute
\bea
[\nabla^{g}_{i}\nabla^{g}_{j}\varphi_1] 
&=&\tr\left\{(\cC^+_{(i}-\cC^-_{(i})(\cC^+_{j)}-\cC^-_{j)})
\right\}.
\eea
Finally, we obtain
\bea
[\nabla^g_{i}\nabla^g_{j} (e^{\omega}\varphi_1)]
&=& 
\left(\omega_{ij}+\omega_{i}\omega_{j} 
\right)\tr I
+\tr\left\{(\cC^+_{(i}-\cC^-_{(i})(\cC^+_{j)}-\cC^-_{j)})
\right\}.
\eea


By collecting the above results we obtain
\bea
b_{0,1}
 &=&
ts\tr\Biggl\{ 
\Biggr[\frac{1}{6}
G^{ij}\left(
R^+_{ij}+R^-_{ij}
-2R^g_{ij}
+W_{ij}
+W_{i}W_{j}\right) 
-G^{ij}G^{kl}\Sigma_{ikl}W_{j}
-\frac{1}{4}G^{ij}G^{kl}\Sigma_{ijkl}
\nonumber\\
&&
+\frac{1}{6}G^{il}G^{jm}G^{kn}\Sigma_{ijk}\Sigma_{lmn}
+\frac{1}{4}G^{ij}G^{lm}G^{kn}\Sigma_{ijk}\Sigma_{lmn}
\Biggr]I
\nonumber\\
&&
+G^{ij}(\cC^+_{i}-\cC^-_{i})(\cC^+_{j}-\cC^-_{j})
\Biggr\}.
\label{720viax}
\eea

Next, we compute the derivatives of the function $\Sigma$
defined by (\ref{529zzc}). By using the eqs. (\ref{361zaa})
and (\ref{362zaa}) we obtain eqs. (\ref{139zax})
and (\ref{140zax}).
By using the results (\ref{72viax}) and (\ref{720viax}) 
we obtain (\ref{134via}),
which proves Theorem \ref{theorem2}; 
the Corollary \ref{corollary2} follows.
It is easy to see that 
for equal operators $L_+=L_-$ the coefficient $B_1$ is equal to 
$(t+s)^{1-n/2}A_1$, as it should.

\subsection{Dirac Type Operators (Proof of Theorem \ref{theorem3}.)}





\subsubsection{Coefficient $c_0$}

The coefficient $c_0$ is given by (\ref{662vian}),
\be
c_0 = c^{(0)}_{0,1}+c^{(0)}_{1,0}+c^{(1)}_{0,0};
\ee
and, since  $c^{(0)}_{1,0}=c^{(1)}_{0,0}=0$ it is equal to 
$c_0= c^{(0)}_{0,1}$,
which is determined by the
second coefficient of the asymptotics of the function 
$Y_0$, (\ref{650via}), which is equal to $Y_0=Y_0^{(0)}$
given by (\ref{642aab}).


First, we 
have
\bea
N^{(0)}_{0} &=& \frac{1}{4}(ts)^{-1}
g^{1/2}(x)G^{1/2}(x')e^{\omega(x,x')}\varphi_2(x,x'),
\label{724via}
\eea
where $\omega$ is defined by (\ref{75via}) and
\be
\varphi_2(x,x')=
\frac{1}{2}
\tr\left\{\mu_+(x,x')\mu_-(x',x)+\mu_-(x,x')\mu_+(x',x)\right\},
\ee
with
\be
\mu_\pm(x,x')=\nu_\pm(x,x')\cP_\pm(x,x')=
\sigma^\pm_{,j}(x,x')i\gamma_\pm^{j}(x)\cP_\pm(x,x').
\ee

We use Lemma \ref{lemma2}, namely, eq. (\ref{533via}) to compute it.
We notice that the diagonal values of the function $\varphi$
and its first derivative vanish,
\be
[\varphi_2]=[\nabla^g_i\varphi_2]=0.
\label{734viax}
\ee
Therefore, by using (\ref{533via}), (\ref{734viax}) and (\ref{631via}) we get
\be
c^{(0)}_{0,1} = \frac{1}{4}
G^{ij}[\nabla^g_i\nabla^g_j(e^\omega\varphi_2)].
\ee

We use now the connection $\nabla^{g,\cA}$ defined with respect to the metric
$g_{ij}(t,s)$ given by
(\ref{113via}) 
and the connection $\cA_i(t,s)$ given by (\ref{114via}).
Now, by using (\ref{639via}) and the diagonal values
of the second derivatives of the Ruse-Synge function $\sigma_\pm$
(\ref{45via}), 
we compute
\bea
[\nabla^{g}_j\nu_\pm(x,x')]=[\nabla^g_{j}\mu_\pm(x,x')]
=-[\nabla^{g}_{j}\nu^\pm(x',x)]
=-[\nabla^g_j\mu_\pm(x',x)]
=ig^\pm_{ji}\gamma_\pm^i.
\label{661via}
\eea

By using (\ref{642via}),
the derivatives of the functions $M_\pm$ and $\cP_\pm$,
(\ref{511mmm}), (\ref{420baa}), and (\ref{661via}), we obtain
\bea
[\nabla^g_i\nabla^g_j (e^\omega\varphi_2)]
=[\nabla^g_i\nabla^g_j \varphi_2]
=2g^+_{k(i}g^-_{j)m}\tr \left(\gamma_+^k\gamma_-^{m}\right),
\label{734via}
\eea
and, therefore, by using (\ref{127via}) we obtain
\be
c_0(t,s)=\frac{1}{2}g_{ij}(t,s)\tr \left(\gamma_+^{i}\gamma_-^{j}\right),
\ee
which gives (\ref{138via}).






\subsubsection{Coefficient $c_1$}

The coefficient $c_1$ is given by (\ref{662vian})
\bea
c_1 &=&  c^{(0)}_{2,0}+ c^{(1)}_{1,0}
+c^{(2)}_{0,0}
+c^{(0)}_{1,1} + c^{(1)}_{0,1} 
+c^{(0)}_{0,2}
\eea
and, since $c^{(1)}_{1,0}=c^{(0)}_{2,0}=0$, is equal to
\bea
c_1
&=&
c^{(2)}_{0,0}
+c^{(0)}_{1,1} + c^{(1)}_{0,1} 
+c^{(0)}_{0,2}.
\eea

The coefficient $c^{(2)}_{0,0}$ is given by (\ref{652via})
\be
c^{(2)}_{0,0}=\tr \left\{[D_+ a^+_0][D_-a^-_0]
\right\},
\ee
and, therefore, by using (\ref{68viab})
we obtain
\be
c^{(2)}_{0,0}=\tr \left(S_+S_-\right).
\label{742viax}
\ee





The coefficient $c^{(0)}_{1,1}$ 
is determined by the second coefficient of the asymptotic
expansion of the integral $Y^{(0)}_{1}$, (\ref{642aab}),
of the function $N^{(0)}_1$, (\ref{642via}), which we can rewrite in the form
\be
N^{(0)}_{1} = \frac{1}{4}(ts)^{-1}
g^{1/2}(x)G^{1/2}(x')e^{\omega(x,x')}\varphi_3(x,x'),
\ee
where
\bea
\varphi_3(x,x') &=& 
-\frac{1}{2}\tr\Biggl\{
t \nu_+(x,x')a_{1}^+(x,x')\mu_-(x',x)
+s \mu_+(x,x')\nu_-(x',x) a_1^-(x',x)
\nonumber\\
&&
+s \nu_-(x,x') a_1^-(x,x')\mu_+(x',x)
+t\mu_-(x,x') \nu_+(x',x)a_{1}^+(x',x)
\Biggr\}.
\eea
We use Lemma {\ref{lemma2}} and eq. (\ref{533via}) to compute it. 
First of all, we notice that
\be
[\varphi_3]=[\nabla^g_i\varphi_3]=0.
\ee
Therefore, we get
\bea
c^{(0)}_{1,1} &=& \frac{1}{4}
G^{ij}[\nabla^g_i\nabla^g_j(e^\omega\varphi_3)].
\eea
Next, by using (\ref{661via})
we compute the diagonal values of the second
derivatives
\be
[\nabla^g_i\nabla^g_j(e^\omega\varphi_3)] 
=[\nabla^g_i\nabla^g_j\varphi_3]
= 2g^+_{m(i}g^-_{j)k}\tr\Biggl\{
t\gamma_-^{k}\gamma_+^{m}
\left(\frac{1}{6}R_+ I-Q_+\right)
+s\gamma_+^{m}\gamma_-^{k}
\left(\frac{1}{6}R_- I-Q_-\right)
\Biggr\}.
\ee
This gives 
\bea
c^{(0)}_{1,1} &=& 
\frac{1}{2}g_{ij}\tr\Biggl\{
t\gamma_-^{j}\gamma_+^{i}
\left(\frac{1}{6}R_+ I-Q_+\right)
+s
\gamma_+^{i}\gamma_-^{j} \left(\frac{1}{6}R_- I-Q_-\right)
\Biggr\}.
\label{748viax}
\eea
Recall that $Q_\pm$ for Dirac type operators is given by
(\ref{qsxxa}).





The coefficient
$c^{(1)}_{0,1}$ 
 is determined by the second coefficient of the asymptotic
expansion of the integral $Y_0^{(1)}$, (\ref{642aab}),  of the function
$N^{(1)}_0$, (\ref{645viax}),
which can be written in the form
\be
N^{(1)}_{0} = \frac{1}{4}(ts)^{-1}g^{1/2}(x)G^{1/2}(x')
e^{\omega(x,x')}
\varphi_4(x,x'),
\ee
where
\bea
\varphi_4(x,x') &=&
\tr\Biggl\{
-t\left(
\theta_+(x,x')\mu_-(x',x)
+ \mu_-(x,x')\theta_+(x',x)\right)
\nonumber\\
&&
-s\left(
\theta_-(x,x')\mu_+(x',x)
+\mu_+(x,x')\theta_-(x',x)
\right)
\Biggr\},
\eea
with
\bea
\theta_\pm(x,x')  &=& e^{-\zeta\pm}D_\pm \left(e^{\zeta_\pm}\cP_\pm\right)
=\left\{i\gamma_\pm^k(x)(\nabla^\pm_k+\zeta^\pm_{,k})+S_\pm(x)\right\}\cP_\pm(x,x'),
\eea

We use Lemma \ref{lemma2} and eq. (\ref{533via}) to compute it.
First of all, we notice that
\be
[\varphi_4]=0.
\ee
Next, by using (\ref{661via})
and the obvious limit
\be
[\theta_\pm] = S_\pm
\label{741via}
\ee
we compute 
\be
[\nabla^g_j\varphi_{4}]=0.
\ee
Therefore,
\bea
c^{(1)}_{0,1} &=&
\frac{1}{4}
G^{ij}[\nabla^g_{i}\nabla^g_{j}(e^\omega\varphi_4)].
\label{753viax}
\eea

Further, by using (\ref{741via}) and (\ref{661via}) 
and omitting all terms that vanish on the diagonal we obtain
\bea
[\nabla^g_i\nabla^g_j(e^\omega\varphi_4)] 
&=& [\nabla^g_i\nabla^g_j\varphi_4]=
-2\tr\Biggl\{\frac{1}{2}t S_+\left[\nabla^{g}_{(i}\nabla^{g}_{j)}\mu_-(x',x)
+\nabla^{g}_{(i}\nabla^{g}_{j)}\mu_-(x,x')
\right]
\nonumber\\
&&
+\frac{1}{2}s S_-\left[\nabla^{g}_{(i}\nabla^{g}_{j)}\mu_+(x',x)
+\nabla^{g}_{(i}\nabla^{g}_{j)}\mu_+(x,x')\right]
\nonumber\\
&&
+t i\gamma_-^kg^-_{k(i}\left[\nabla^g_{j)}\theta_+(x',x)
-\nabla^{g}_{j)}\theta_+(x,x')
\right]
\nonumber\\
&&
+s i\gamma_+^kg^+_{k(i}\left[\nabla^g_{j)}\theta_-(x',x)
-\nabla^{g}_{j)}\theta_-(x,x')
\right]
\Biggr\}.
\label{754viax}
\eea

Next, by using (\ref{716viax}) and (\ref{661via}) we compute
\bea
[\nabla^g_{(i}\nabla^g_{j)}\mu_\pm(x,x')] &=& [\nabla^g_{(i}\nabla^g_{j)}\nu_\pm(x,x')]
-2i\gamma^k_\pm g^\pm_{k(i}\cA^\pm_{j)},
\\
{}[\nabla^g_{(i}\nabla^g_{j)}\mu_\pm(x',x)] &=& [\nabla^g_{(i}\nabla^g_{j)}\nu_\pm(x',x)]
-2i\gamma^k_\pm g^\pm_{k(i}\cA^\pm_{j)}.
\eea
Now, by using (\ref{310viax}) and (\ref{360mmm}) we 
have
\be
\nabla^{g}_i\gamma_\pm^k
=-W_\pm{}^k{}_{im}\gamma_\pm^m
-[\cA^\pm_i,\gamma_\pm^k]
\label{758viax}
\ee
By using this equation and (\ref{361zaa}) we compute
\bea
[\nabla^{g}_{(i}\nabla^{g}_{j)}\nu_\pm(x,x')]
=-2g^\pm_{k(i}i
[\cA^\pm_{j)},\gamma_+^{k}]
+i\gamma_\pm^k g^\pm_{mk}W_\pm{}^m{}_{ij}.
\label{763via}
\eea
Further, by using (\ref{467viax})
we have
\bea
{}[\nabla^{g}_{(i}\nabla^{g}_{j)}\nu_\pm(x',x)]
=-i\gamma_\pm^kg^\pm_{mk}W_\pm{}^m{}_{ij}.
\label{764via}
\eea
Therefore,
\bea
[\nabla^g_{(i}\nabla^g_{j)}\mu_\pm(x,x')] &=& 
i\gamma_\pm^k g^\pm_{mk}W_\pm{}^m{}_{ij}
-2\cA^\pm_{j)}g^\pm_{i)k}i\gamma_\pm^{k},
\nonumber\\
\label{765via}\\
{}[\nabla^g_{(i}\nabla^g_{j)}\mu_\pm(x',x)] &=& 
-i\gamma_\pm^kg^\pm_{mk}W_\pm{}^m{}_{ij}
-2i\gamma^k_\pm g^\pm_{k(i}\cA^\pm_{j)}.
\label{766via}
\eea

Next, by using (\ref{512mmm}) and (\ref{422qqc}) we compute 
the diagonal values
\bea
[\nabla^+_j\theta_+(x,x')] &=& 
i\gamma_+^k\Omega^+_{jk}+\nabla_j^+S^+,
\\{}
[\nabla^-_j\theta_-(x',x)] &=& 
-i\gamma_-^k\Omega^-_{jk},
\eea
where
\be
\Omega^\pm_{jk}=\frac{1}{2}\cR^\pm_{jk}
+\frac{1}{6}R^\pm_{jk}I.
\label{768viax}
\ee
To avoid confusion, we note that the derivatives $\nabla^g$ 
here do not include the connection $\cA_\pm$;
therefore,
\bea
[\nabla^{g}_j\theta_\pm(x,x')] &=& 
i\gamma_\pm^k\Omega^\pm_{jk}+\nabla_j^\pm S^\pm-\cA^\pm_jS_\pm,
\\{}
[\nabla^{g}_j\theta_\pm(x',x)] &=& 
-i\gamma_\pm^k\Omega^\pm_{jk}+S_\pm\cA^\pm_j.
\eea

By using the above results we compute
\bea
[\nabla^g_i\nabla^g_j\varphi_4] &=&
-2\tr\Biggl\{
-t i\gamma_-^mg^-_{m(i}\nabla_{j)}^+ S_+
+t S_+ \left[
\left(\cC^+_{(j}-\cC^-_{(j}\right)g^-_{i)k}i\gamma_-^{k}
+i\gamma_-^{k}g^-_{k(i}\left(\cC^+_{j)}-\cC^-_{j)}\right)
\right]
\nonumber\\
&&
-s i\gamma_+^mg^+_{m(i}\nabla_{j)}^- S_-
-s S_- \left[
\left(\cC^+_{(j}-\cC^-_{(j}\right)g^+_{i)k}i\gamma_+^{k}
+i\gamma_+^{k}g^+_{k(i}\left(\cC^+_{j)}-\cC^-_{j)}\right)
\right]
\nonumber\\
&&
+2t \gamma_-^m \gamma_+^kg^-_{m(i}\Omega^+_{j)k}
+2s \gamma_+^m \gamma_-^kg^+_{m(i}\Omega^-_{j)k}
\Biggr\}.
\eea
Now, by using (\ref{312viax}) and the cyclicity of the trace we obtain a simpler
form
\bea
[\nabla^g_i\nabla^g_j\varphi_4] &=&
-2\tr\Biggl\{
-t i\gamma_-^mg^-_{m(i}\nabla_{j)}^+ S_+
-s i\gamma_+^mg^+_{m(i}\nabla_{j)}^- S_-
\nonumber\\
&&
+2t \gamma_-^m \gamma_+^kg^-_{m(i}\Omega^+_{j)k}
+2s \gamma_+^m \gamma_-^kg^+_{m(i}\Omega^-_{j)k}
\Biggr\}.
\eea

Thus, we obtain the coefficient $c^{(1)}_{0,1}$ 
from (\ref{753viax})
\bea
c^{(1)}_{0,1} &=&
-\frac{1}{2}
G^{ij}\tr\Biggl\{
-t i\gamma_-^mg^-_{m(i}\nabla_{j)}^+ S_+
-s i\gamma_+^mg^+_{m(i}\nabla_{j)}^- S_-
\nonumber\\
&&
+2t \gamma_-^m \gamma_+^kg^-_{m(i}\Omega^+_{j)k}
+2s \gamma_+^m \gamma_-^kg^+_{m(i}\Omega^-_{j)k}
\Biggr\}.
\label{771viax}
\eea


The coefficient $c^{(0)}_{0,2}$ is determined by the third
coefficient of the asymptotic expansion of the integral
$Y^{(0)}_0$, (\ref{642aab}), of the function $N^{(0)}_0$, (\ref{724via}).
We use Lemma \ref{lemma2} to compute it.
Since the function $\varphi_2$ and its first derivative vanish on the diagonal,
(\ref{734viax}),
it is given by the eq. (\ref{533viax}).
We use the equations
\bea
[\nabla^g_{(i}\nabla^g_j\nabla^g_k\nabla^g_{l)}(e^\omega\varphi_2)]
&=&
[\nabla^g_{(i}\nabla^g_j\nabla^g_k\nabla^g_{l)}\varphi_2]
+4\omega_{(i}[\nabla^g_{j}\nabla^g_k\nabla^g_{l)}\varphi_2]
\nonumber\\
&&
+6\left(\omega_{(ij}+\omega_{(i}\omega_j\right)[\nabla^g_{k}\nabla^g_{l)}\varphi_2],
\\{}
[\nabla^g_{(j}\nabla^g_k\nabla^g_{l)}(e^\omega\varphi_2)]
&=&
[\nabla^g_{(j}\nabla^g_k\nabla^g_{l)}\varphi_2]
+3\omega_{(j}[\nabla^g_{k}\nabla^g_{l)}\varphi_2],
\\{}
[\nabla^g_{(k}\nabla^g_{l)}(e^\omega\varphi_2)]
&=&
[\nabla^g_{(k}\nabla^g_{l)}\varphi_2],
\eea
and (\ref{711viaxz}) to obtain
\bea
c^{(0)}_{0,2} &=& 
\frac{1}{4}ts
\Biggl\{
\frac{1}{2}G^{ij}G^{kl}
[\nabla^g_{(i}\nabla^g_j\nabla^g_k\nabla^g_{l)}\varphi_2]
+N^{jkl}[\nabla^g_{(j}\nabla^g_k\nabla^g_{l)}\varphi_2]
\label{774viax}\\
&&
+\left[
\frac{1}{6}\left(G^{kl}G^{ij}
+2G^{ik}G^{jl}\right)
\left(R^+_{ij}+R^-_{ij}-2R^g_{ij}\right)
+M^{kl}
\right]
[\nabla^g_{(k}\nabla^g_{l)}\varphi_2]
\Biggr\},
\nonumber
\eea
where $N^{jkl}$ and $M^{kl}$ are defined by
(\ref{144viax}) and (\ref{147zax}).


The second derivative of $\varphi_2$ was computed in (\ref{734via}).
So, we compute the third derivative; we have
\bea
[\nabla^g_{(i}\nabla^g_j\nabla^g_{k)}\varphi_2]
&=&
\frac{3}{2}\Biggl[\nabla^g_{(i}\nabla^g_j\mu_+(x,x')\nabla^g_{k)}\mu_-(x',x)
+\nabla^g_{(i}\mu_+(x,x')\nabla^g_{j}\nabla^g_{k)}\mu_-(x',x)
\nonumber\\
&&
+\nabla^g_{(i}\nabla^g_j\mu_-(x,x')\nabla^g_{k)}\mu_+(x',x)
+\nabla^g_{(i}\mu_-(x,x')\nabla^g_{j}\nabla^g_{k)}\mu_+(x',x)
\Biggr].
\nonumber\\
\eea
By using (\ref{661via}), (\ref{765via}) and (\ref{766via}) we get
\bea
[\nabla^g_{(i}\nabla^g_j\nabla^g_{k)}\varphi_2]
&=&
3\tr\Biggl\{
\left[W_+{}^m{}_{(ij}g^-_{k)q}g^+_{mp}
+W_-{}^m{}_{(ij} g^+_{k)p} g^-_{mq}\right]\gamma_+^p\gamma_-^{q}
\label{779viax}
\\
&&
+g^-_{q(i}(\cC^+_{j}-\cC^-_j) g^+_{k)p}\gamma_-^{q}\gamma_+^p
-g^+_{p(i}(\cC^+_{j}-\cC^-_j)g^-_{k)q}\gamma_+^p\gamma_-^{q}
\nonumber
\Biggr\}.
\eea


Finally, we compute the forth derivative of the function $\varphi_2$;
we rewrite it in the form
\be
\varphi_{2}=-\frac{1}{2}A_{pq'}(x,x')B^{pq'}(x,x')
+(+\leftrightarrow -),
\ee
where
\bea
A_{pq'} &=& \sigma^+_{,p}\sigma^-_{,q'},
\\
B^{pq'} &=& \tr\left\{
\cP_-^{-1}(x,x')\gamma^p_+(x)\cP_+(x,x')\gamma_-^{q'}(x')
\right\},
\eea
and the symbol $(+\leftrightarrow -)$ indicates that one should add
the same term with $+$ and $-$ switched.
The diagonal value of the forth symmetrized derivative is then
\bea
[\nabla^g_{(i}\nabla^g_j\nabla^g_{k}\nabla^g_{l)}\varphi_2] &=&
-\frac{1}{2}
[\nabla^g_{(i}\nabla^g_j\nabla^g_{k}\nabla^g_{l)}A_{pq'}]
\tr\left(\gamma_+^p\gamma_-^q\right)
-2[\nabla^g_{(i}\nabla^g_j\nabla^g_{k}A_{pq'}][\nabla^g_{l)}B^{pq'}]
\nonumber\\
&&
-3[\nabla^g_{(i}\nabla^g_jA_{pq'}]
[\nabla^g_{k}\nabla^g_{l)}B^{pq'}]
+(+\leftrightarrow -).
\eea

First of all, it is easy to get
\bea
[\nabla^g_{(i}\nabla^g_{j)}A_{pq'}] &=& -2g^+_{p(i}g^-_{j)q},
\\{}
[\nabla^g_{(i}\nabla^g_j\nabla^g_{k)}A_{pq'}] &=&
3V^-_{q(ij}g^+_{k)p}
-3T^+_{p(ij}g^-_{k)q},
\eea
where the tensors $V_{ijk}$ and $T_{ijk}$ are given by
(\ref{467viax}), (\ref{361zaa})
\bea
T^\pm_{ijk} &=& 3g^\pm_{m(i}W_\pm^m{}_{jk)},
\\{}
V^\pm_{ijk} &=& -g^\pm_{mi}W_\pm^m{}_{kj}.
\eea
Similarly, we obtain
\bea
[\nabla^g_{(i}\nabla^g_j\nabla^g_{k}\nabla^g_{l)}A_{pq'}]
&=& 4 V^-_{q(ijk}g^+_{l)p}
+6 V^-_{q(ij}T^+_{kl)p}
-4 T^+_{p(ijk}g^-_{l)q},
\eea
where the tensors $T^\pm_{ijkl}$ and $V^\pm_{ijkl}$ are given by
(\ref{469viax}), (\ref{470viax}),
\bea
T^\pm_{ijkl}
&=&
3g^\pm_{m(j} \nabla^g{}_{k}W_\pm^m{}_{l)i}
+g^\pm_{mi} \nabla^g{}_{(j}W_\pm^m{}_{kl)}
+3g^\pm_{m(j}W_\pm^n{}_{k|i|}W_\pm^{m}{}_{l)n}
\nonumber\\
&&
+g^\pm_{mi}W_\pm^{n}{}_{(jk}W_\pm^{m}{}_{l)n}
+3g^\pm_{nm}W_\pm^n{}_{(jk}W_\pm^{m}{}_{l)i},
\label{469viaxz}\\
V^\pm_{ijkl}
&=& -g^\pm_{mi} \nabla^g{}_{(j}W_\pm^m{}_{kl)}
-g^\pm_{mi}W_\pm^n{}_{(jk}W_\pm^{m}{}_{l)n}.
\label{470viaxz}
\eea

Next, by using
\be
[\nabla_l^+B^{pq'}]
=-\tr\left\{(\cC^+_l-\cC^-_l)\gamma_+^p\gamma_-^q\right\}
\ee
we compute
\be
[\nabla^g_lB^{pq'}]
=\tr\left\{
-(\cC^+_l-\cC^-_l)\gamma_+^p\gamma_-^q
-W_+{}^p{}_{lm}\gamma_+^m\gamma_-^q
\right\}.
\ee
Further, we have
\bea
[\nabla^g_{(k}\nabla^g_{l)}B^{pq'}] &=& 
\tr\Biggl\{\left[-\nabla^g_{(k}W_+{}^p{}_{l)m}
+3W_+{}^p{}_{n(k}W_+{}^n{}_{l)m}
-W_+{}^n{}_{kl}W_+{}^p{}_{nm}\right]
\gamma_+^m\gamma_-^q
\nonumber\\
&&
+2W_+{}^p{}_{m(k}(\cC^+_{l)}-\cC^-_{l)})\gamma_+^m\gamma_-^q
-W_+{}^n{}_{kl}(\cC^+_n-\cC^-_n)\gamma_+^p\gamma_-^q
\Biggr\}
\nonumber\\
&&
+[\nabla^+_{(k}\nabla^{+}_{l)}B^{pq'}].
\eea
Next, by using the equations 
$[\nabla^+_{i}\cP_+]=[\nabla^+_{(i}\nabla^+_{j)}\cP_+]=0$,
$\nabla^+_i\gamma_+^j=0$, and
\bea
[\nabla^+_{i}\cP_-] &=& \cC^+_i-\cC^-_i,
\label{372zaaz}
\\
{}[\nabla^+_{(i}\nabla^+_{j)}\cP_-] &=& 
\nabla^+_{(i}(\cC^+_{j)}-\cC^-_{j)})
+(\cC^+_{(i}-\cC^-_{(i})(\cC^+_{j)}-\cC^-_{j)}),
\eea
we obtain
\bea
[\nabla^+_{(i}\nabla^{+}_{j)}B^{pq'}] &=&
\tr\left\{\left[
-\nabla^+_{(i}(\cC^+_{j)}-\cC^-_{j)})
+(\cC^+_{(i}-\cC^-_{(i})(\cC^+_{j)}-\cC^-_{j)})
\right]\gamma_+^p\gamma_-^q\right\}
\nonumber\\
&=&
\tr\Biggl\{\Bigl[
-\nabla^{g,\cA}_{(i}(\cC^+_{j)}-\cC^-_{j)})
+W_+^s{}_{ij}(\cC^+_{s}-\cC^-_{s})
\nonumber\\
&&
-2\cC^+_{(i}\cC^-_{j)}
+\cC^+_{(i}\cC^+_{j)}+\cC^-_{(i}\cC^-_{j)}
\Bigr]\gamma_+^p\gamma_-^q\Biggr\},
\eea
and, therefore,
\bea
[\nabla^g_{(k}\nabla^g_{l)}B^{pq'}] 
&=&
\tr\Biggl\{
\Bigl[-\nabla^g_{(k}W_+{}^p{}_{l)m}
+3W_+{}^p{}_{n(k}W_+{}^n{}_{l)m}
-W_+{}^n{}_{kl}W_+{}^p{}_{nm}
\Bigr]
\gamma_+^m\gamma_-^q
\nonumber\\
&&
+2W_+{}^p{}_{m(k}(\cC^+_{l)}-\cC^-_{l)})\gamma_+^m\gamma_-^q
-\nabla^{g,\cA}_{(k}(\cC^+_{l)}-\cC^-_{l)})\gamma_+^p\gamma_-^q
\nonumber\\
&&
+\left(\cC^+_{(k}\cC^+_{l)}-2\cC^+_{(k}\cC^-_{l)}+\cC^-_{(k}\cC^-_{l)}\right)
\gamma_+^p\gamma_-^q
\Biggr\}.
\eea

Finally, by collecting all these results we obtain
\bea
[\nabla^g_{(i}\nabla^g_j\nabla^g_{k}\nabla^g_{l)}\varphi_2] 
&=&
\mathrm{Sym}(i,j,k,l)
\tr\Biggl\{
V_{pqijkl}\gamma_+^p\gamma_-^q
%
%
-6g^+_{jp}g^-_{qi}
[\gamma_+^p,\gamma_-^q]
\nabla^{g,\cA}_{k}(\cC^+_{l}-\cC^-_{l})
\nonumber\\
&&
%
%
-6\left(
g^+_{mp}W_+^m{}_{ij}g^-_{kq}
+g^+_{kp}g^-_{mq}W_-^m{}_{ij}
\right)
[\gamma_+^p,\gamma_-^q](\cC^+_{l}-\cC^-_{l})
\nonumber\\
&&
+6g^+_{pi}g^-_{jq}\left(\cC^+_{k}\cC^+_{l}+\cC^-_{k}\cC^-_{l}\right)
\left(
\gamma_+^p\gamma_-^q
+\gamma_-^q\gamma_+^p 
\right)
\nonumber\\
&&
-12g^+_{pi}g^-_{jq}\cC^+_{k}\cC^-_{l}\gamma_+^p\gamma_-^q
-12g^+_{pi}g^-_{jq}\cC^-_{l}\cC^+_{k}\gamma_-^q\gamma_+^p
\Biggr\},
\eea
where $V_{pqijkl}$ is defined by (\ref{145viaxz}).

%
%

This enables us to compute the coefficient $c^{(0)}_{0,2}$,
(\ref{774viax}),
\bea
c^{(0)}_{0,2} &=& 
\frac{1}{4}ts\;
\tr\Biggl\{
\frac{1}{3}\left(G^{kl}G^{ij}
+2G^{ik}G^{jl}\right)
\left(R^+_{ij}+R^-_{ij}-2R^g_{ij}\right)
g^+_{p(k}g^-_{l)q}\gamma_+^{p}\gamma_-^q
\nonumber\\
&&
+\Biggl[
\frac{1}{2}G^{(ij}G^{kl)}V_{pqijkl}
+3N^{jkl}
\left(g^+_{mp}W_+{}^m{}_{(jk}g^-_{l)q}
+g^-_{mq}W_-{}^m{}_{(jk} g^+_{l)p}
\right) 
\Biggr]
\gamma_+^{p}\gamma_-^q
\nonumber\\
&&
+2M^{kl}g^+_{p(k}g^-_{l)q}\gamma_+^{p}\gamma_-^q
+3G^{(ij}G^{kl)}g^+_{jp}g^-_{qi}
\left(
\gamma_-^q\gamma_+^p
-\gamma_+^p\gamma_-^q 
\right)
\nabla^{g,\cA}_{k}(\cC^+_{l}-\cC^-_{l})
\nonumber\\
&&
%
%
+\Bigl[3G^{(ij}G^{kl)}\left(
g^+_{mp}W_+^m{}_{ij}g^-_{kq}
+g^+_{kp}g^-_{mq}W_-^m{}_{ij}
\right)
+3N^{jkl}g^+_{pk}g^-_{jq} 
\Bigr]
\nonumber\\
&&
\times
\left(\gamma_-^{q}\gamma_+^p
-\gamma_+^p\gamma_-^{q}
\right)(\cC^+_{l}-\cC^-_{l})
\nonumber\\
&&
+3G^{(ij}G^{kl)}\left(\cC^+_{k}\cC^+_{l}+\cC^-_{k}\cC^-_{l}\right)
\left(
\gamma_+^p\gamma_-^q
+\gamma_-^q\gamma_+^p 
\right)g^+_{pi}g^-_{jq}
\nonumber\\
&&
-6G^{(ij}G^{kl)}\cC^+_{k}\cC^-_{l}\gamma_+^p\gamma_-^qg^+_{pi}g^-_{jq}
-6G^{(ij}G^{kl)}\cC^-_{l}\cC^+_{k}\gamma_-^q\gamma_+^pg^+_{pi}g^-_{jq}
\Biggr\}.
\eea
Thus (after some tedious but straightforward manipulations;
by using (\ref{qsxxa}) and many
well known algebraic properties of the Dirac matrices
\cite{zhelnorovich19})
we obtain the coefficient $c_1$, (\ref{150zax}).
This proves Theorem \ref{theorem3}; the Corollary \ref{corollary3} follows.

For equal operators $D_-=D_+$ the coefficient $C_1$ takes the form
\be
C_1=(t+s)^{-n/2}\int_M dx g^{1/2}\tr\Biggl\{
\left(\frac{n}{2}-1\right)
\left(\frac{1}{6}R+\frac{1}{2}\gamma^{ij}\cR_{ij}-S^2\right)
-\frac{(n-1)}{2}i\gamma^j\nabla_jS
\Biggr\}.
\ee
Notice that since $\nabla_j S$ anticommutes with $\gamma^j$
the last term here vanishes and $C_1$ is indeed equal to
$(n/2-1)(t+s)^{-n/2}A_1$, with $A_1$ given by (\ref{126viax}).


\section{Conclusion}
\setcounter{equation}0

The primary goal of this paper was to introduce and to study some new 
spectral invariants 
of a pair of elliptic partial differential operators on manifolds, 
that we call the relative spectral invariants and the combined heat
traces. Of special interest are the asymptotics of these 
invariants. We established a general asymptotic expansion 
of these invariants and computed the first two coefficients of the
asymptotic expansions.


\end{document}